%% file: INTEGRAL-reloaded.tex
\documentclass[final,5p,times,one,longtitle]{elsarticle}
\usepackage[normalem]{ulem}
\usepackage{lineno,hyperref}
\usepackage{amssymb}
\modulolinenumbers[5]

\journal{New Astronomy Reviews}

\biboptions{authoryear}

\input{aa_macros.tex}

\begin{document}

\begin{frontmatter}

\title{INTEGRAL reloaded: spacecraft, instruments and ground system}

\author[1]{Erik Kuulkers \fnref{myfootnote}}
\ead{Erik.Kuulkers@esa.int}
\ead[url]{www.esa.int}

\author[2]{Carlo Ferrigno}
\author[3]{Peter Kretschmar}
\author[4]{Julia Alfonso-Garz\'on}
\author[5]{Marius Baab}
\author[6]{Angela Bazzano}
\author[3]{Guillaume B\'elanger}
\author[7]{Ian Benson}
\author[8]{Anthony J.\ Bird}
\author[2]{Enrico Bozzo}
\author[9]{Søren Brandt}
\author[5]{Elliott Coe}
\author[3]{Isabel Caballero}
\author[10]{Floriane Cangemi}
\author[9]{Jérôme Chenevez}
\author[11,12]{Bradley Cenko}
\author[5]{Nebil Cinar}
\author[13]{Alexis Coleiro}
\author[7]{Stefano De Padova}
\author[14]{Roland Diehl}
\author[15]{Claudia Dietze}
\author[4]{Albert Domingo}
\author[5]{Mark Drapes}
\author[5]{Eleonora D'uva}
\author[3]{Matthias Ehle}
\author[3]{Jacobo Ebrero}
\author[5]{Mithrajith Edirimanne}
\author[16]{Natan A.\ Eismont}
\author[5]{Timothy Finn}
\author[6]{Mariateresa Fiocchi}
\author[17]{Elena Garcia Tomas}
\author[18]{Gianluca Gaudenzi}
\author[19]{Thomas Godard}
\author[20,21]{Andrea Goldwurm}
\author[10]{Diego Götz}
\author[10]{Christian Gouiffès}
\author[16]{Sergei A.\ Grebenev}
\author[14]{Jochen Greiner}
\author[10]{Aleksandra Gros}
\author[22]{Lorraine Hanlon}
\author[23,24]{Wim Hermsen}
\author[3]{Cristina Hernández}
\author[25]{Margarita Hernanz}
\author[26]{Jutta Huebner}
\author[27,28]{Elisabeth Jourdain}
\author[29]{Giovanni La Rosa}
\author[30]{Claudio Labanti}
\author[13]{Philippe Laurent}
\author[5]{Alexander Lehanka}
\author[9]{Niels Lund}
\author[5]{James Madison}
\author[27,28]{Julien Malzac}
\author[5]{Jim Martin}
\author[4]{J. Miguel Mas-Hesse}
\author[22]{Brian McBreen}
\author[31]{Alastair McDonald}
\author[11]{Julie McEnery}
\author[32]{Sandro Meregehtti}
\author[6]{Lorenzo Natalucci}
\author[3]{Jan-Uwe Ness}
\author[9]{Carol Anne Oxborrow}
\author[5]{John Palmer}
\author[26]{Sibylle Peschke}
\author[33]{Francesco Petrucciani}
\author[34]{Norbert Pfeil}
\author[5]{Michael Reichenbaecher}
\author[6]{James Rodi}
\author[10]{Jérôme Rodriguez}
\author[27,28]{Jean-Pierre Roques}
\author[3]{Emilio Salazar Doñate}
\author[5]{Dave Salt}
\author[3]{Celia Sánchez-Fernández}
\author[10]{Aymeric Sauvageon}
\author[2]{Volodymyr Savchenko}
\author[16]{Sergey Yu.\ Sazonov}
\author[26]{Stefano Scaglioni}
\author[3]{Norbert Schartel}
\author[35]{Thomas Siegert}
\author[26]{Richard Southworth}
\author[16]{Rashid A.\ Sunyaev}
\author[5]{Liviu Toma}
\author[6]{Pietro Ubertini}
\author[24]{Ed P.J.\ van den Heuvel}
\author[14]{Andreas von Kienlin}
\author[5]{Nikolai von Krusenstiern}
\author[1]{Christoph Winkler}
\author[36]{Hajdas Wojciech}
\author[6]{Ugo Zannoni}

\address[1]{European Space Agency (ESA), European Space Research and Technology Centre (ESTEC), Keplerlaan 1, 2201 AZ Noordwijk, The Netherlands}
\address[2]{ISDC/University of Geneva, Chemin d'\'Ecogia 16, 1290 Versoix, Switzerland}
\address[3]{European Space Agency (ESA), European Space Astronomy Centre (ESAC), Camino Bajo del Castillo s/n, 28692 Villanueva de la Ca\~{n}ada, Madrid, Spain}
\address[4]{Centro de Astrobiolog\'ia (CSIC--INTA),28692 Villanueva de la Ca\~nada, Madrid, Spain}
\address[5]{TPZ Vega for the European Space Agency (ESA), European Space Operations Centre (ESOC), Robert-Bosch-Stra{\ss}e 5, 64293 Darmstadt, Germany}
\address[6]{INAF, IAPS, Via Fosso del Cavaliere 100, 00133-Rome, Italy}
\address[7]{SERCO for the European Space Agency (ESA), European Space Operations Centre (ESOC), Robert-Bosch-Stra{\ss}e 5, 64293 Darmstadt, Germany}
\address[8]{Department of Physics \& Astronomy, University of Southampton, Highfield, Southampton, SO17 1BJ, UK}
\address[9]{DTU Space—National Space Institute, Technical University of Denmark, Elektrovej 327-328, DK-2800 Lyngby, Denmark}
\address[10]{Laboratoire AIM, CEA/CNRS/Université Paris-Saclay, Université de Paris, Orme des Merisiers, 91191 Gif-sur-Yvette, France}
\address[11]{Astrophysics Science Division, NASA Goddard Space Flight Center, Mail Code 661, Greenbelt, MD 20771, USA}
\address[12]{Joint Space-Science Institute, University of Maryland, College Park, MD 20742, USA}
\address[13]{Lab.\ Astroparticule et Cosmologie, CNRS, Université de Paris, CEA, Observatoire de Paris, 10 rue Alice Domon et Léonie Duquet, 75013, Paris, France}
\address[14]{Max Planck Institut f\"ur extraterrestrische Physik, D-85748 Garching, Germany}
\address[15]{CS for the European Space Agency (ESA), European Space Operations Centre (ESOC), Robert-Bosch-Stra{\ss}e 5, 64293 Darmstadt, Germany}
\address[16]{Space Research Institute, Russian Academy of Science, Profsoyuznaya 84/32, 117997 Moscow, Russia}
\address[17]{LSE Space for the European Space Agency (ESA), European Space Operations Centre (ESOC), Robert-Bosch-Stra{\ss}e 5, 64293 Darmstadt, Germany}
\address[18]{SCISYS for the European Space Agency (ESA), European Space Operations Centre (ESOC), Robert-Bosch-Stra{\ss}e 5, 64293 Darmstadt, Germany}
\address[19]{Rhea for the European Space Agency (ESA), European Space Operations Centre (ESOC), Robert-Bosch-Stra{\ss}e 5, 64293 Darmstadt, Germany}
\address[20]{Université de Paris, CNRS, AstroParticule et Cosmologie, F-75013, Paris, France}
\address[21]{Département d’Astrophysique, IRFU / CEA Saclay, F-91191 Gif-sur-Yvette, France}
\address[22]{School of Physics \&\ Centre for Space Research, University College Dublin, Belfield, Dublin 4, Ireland}
\address[23]{SRON Netherlands Institute for Space Research, Sorbonnelaan 2, NL-3584 CA Utrecht, the Netherlands}
\address[24]{Anton Pannekoek Institute for Astronomy, University of Amsterdam, Science Park 904, NL-1098 XH Amsterdam, the Netherlands}
\address[25]{Institut de Ciències de l’Espai (CSIC-IEEC), Campus UAB, Facultat de Ciències, C5 parell 2$^{\rm on}$, 08193 Bellaterra, Barcelona, Spain}
\address[26]{European Space Agency (ESA), European Space Operations Centre (ESOC), Robert-Bosch-Stra{\ss}e 5, 64293 Darmstadt, Germany}
\address[27]{CNRS, Institut de Recherche Astrophysique et Plan\'etologie (IRAP), 9 Av. colonel Roche, BP 44346, F-31028 Toulouse cedex 4, France}
\address[28]{Universit\'e de Toulouse, UPS-OMP, IRAP, Toulouse, France}
\address[29]{INAF, IASF-Palermo, Via Ugo La Malfa 153, 90146-Palermo, Italy}
\address[30]{INAF - OAS, Via P.\ Gobetti 101, 40129 Bologna, Italy}
\address[31]{CGI for the European Space Agency (ESA), European Space Operations Centre (ESOC), Robert-Bosch-Stra{\ss}e 5, 64293 Darmstadt, Germany}
\address[32]{INAF, IASF-Milano, via A. Corti 12, I-20133 Milano, Italy}
\address[33]{CSSI for the European Space Agency (ESA), European Space Operations Centre (ESOC), Robert-Bosch-Stra{\ss}e 5, 64293 Darmstadt, Germany}
\address[34]{TERMA for the European Space Agency (ESA), European Space Operations Centre (ESOC), Robert-Bosch-Stra{\ss}e 5, 64293 Darmstadt, Germany}
\address[35]{Center for Astrophysics and Space Sciences, University of California, San Diego, 9500 Gilman Dr, 92093-0424, La Jolla, USA}
\address[36]{Paul Scherrer Institute (PSI), Forschungsstrasse 111, 5232 Villigen, Switzerland}





\begin{abstract}
The European Space Agency's INTErnational Gamma-Ray Astrophysics Laboratory (ESA/INTEGRAL) was launched aboard a Proton-DM2 rocket on 17 October 2002 at 06:41 CEST, from Baikonur in Kazakhstan. Since then, INTEGRAL has been providing long, uninterrupted observations (up to about 47\,hr, or 170\,ksec, per satellite orbit of 2.7~days) with a large field-of-view (FOV, fully coded: 100 deg$^2$), millisecond time resolution, keV energy resolution, polarization measurements, as well as additional wavelength coverage at optical wavelengths.
This is realized by two main instruments in the 15\,keV to 10\,MeV energy range, the spectrometer SPI (spectral resolution 3\,keV at 1.8\,MeV) and the imager IBIS (angular resolution: 12\,arcmin FWHM),
complemented by X~ray (JEM-X; 3--35\,keV) and optical (OMC; Johnson V-band) monitor instruments.  All instruments are co-aligned to simultaneously observe the target region.
A particle radiation monitor (IREM) measures charged particle fluxes near the spacecraft.
The Anti-coincidence subsystems of the main instruments, built to reduce the background, are also very efficient all-sky $\gamma$-ray detectors, which provide virtually omni-directional monitoring above $\sim$75\,keV.
Besides the long, scheduled observations, INTEGRAL can rapidly (within a couple of hours) re-point and conduct Target of Opportunity (ToO) observations on a large variety of sources.

INTEGRAL observations and their scientific results have been building an impressive legacy: The discovery of currently more than 600 new high-energy sources; the first-ever direct detection of $^{56}$Ni and $^{56}$Co radio-active decay lines from a Type Ia supernova; spectroscopy of isotopes from galactic nucleo-synthesis sources; new insights on enigmatic positron annihilation in the Galactic bulge and disk; and  pioneering gamma-ray polarization studies. INTEGRAL is also a successful actor in the new multi-messenger astronomy introduced by non-electromagnetic signals from gravitational waves and from neutrinos: INTEGRAL found the first prompt
electromagnetic radiation in coincidence with a binary neutron star merger.
Up to now more than 1750 scientific papers based on INTEGRAL data have been published in refereed journals. In this paper, we will give a comprehensive update of the satellite status
after more than 18 years of operations in a harsh space environment, and an account of the successful Ground Segment.

\end{abstract}

\begin{keyword}
\texttt{Gamma-ray Observatory\sep INTEGRAL \sep Gamma-ray instruments \sep Gamma-ray sources}
\MSC[2010] 00-01\sep  99-00
\end{keyword}

\end{frontmatter}

\newpageafter{author}

\section{INTErnational Gamma-Ray Astrophysics Laboratory - INTEGRAL}

\begin{figure*}[ht!]
  \centering
  \includegraphics[width=0.8\linewidth]{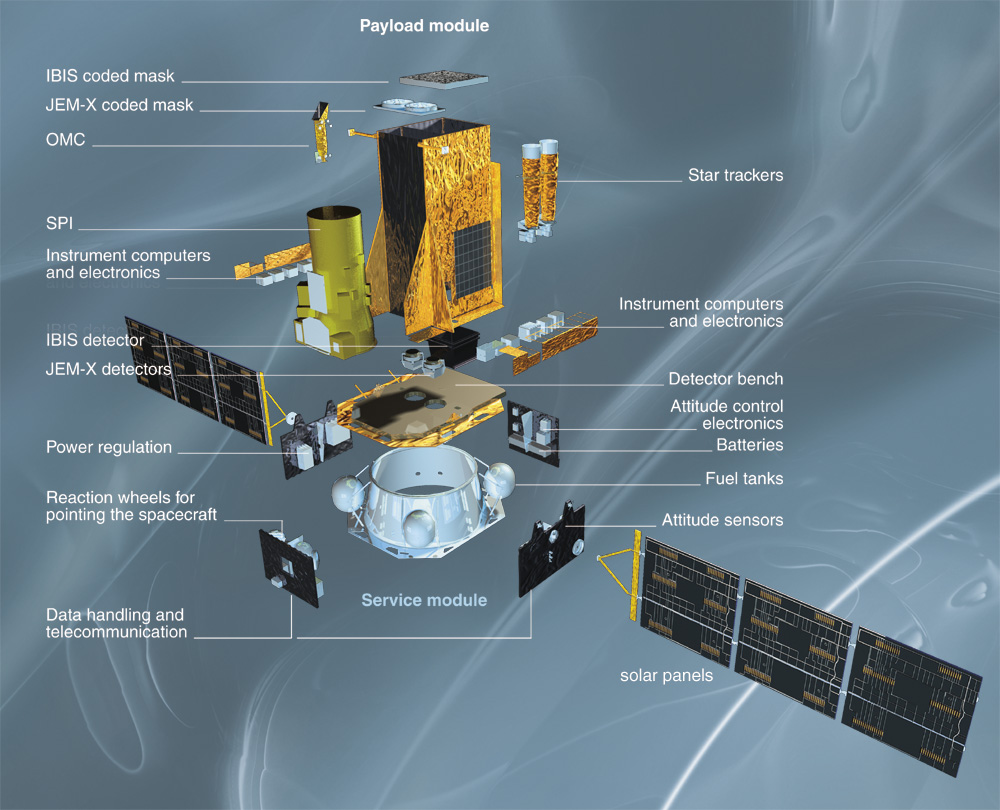}
  \caption{Diagram of the INTEGRAL spacecraft and its instruments.}
  \label{fig:integral}
\end{figure*}

INTEGRAL \citep{Winkler2003} carries onboard a variety of scientific instruments (see Fig.~\ref{fig:integral}):
the `SPectrometer aboard INTEGRAL' (SPI), the `Imager on Board the INTEGRAL Satellite' (IBIS), two
`Joint European X-Ray Monitors' (JEM-X1 and JEM-X2) and the `Optical Monitoring Camera' (OMC).
JEM-X, IBIS and SPI have coded masks which realize a wide field-of-view (FoV) and operate simultaneously in the 15\,keV to 10\,MeV energy range
(JEM-X: 3--35\,keV; IBIS: 15\,keV--10\,MeV; SPI: 18\,keV--8\,MeV).
SPI is ideal for high-resolution spectrometry in the hard X-ray and gamma-ray range extending to nuclear lines (i.e., MeV energies), as expected
from, e.g., radioactive isotopes and their decay.
IBIS main capability is imaging in the hard X-ray and gamma-ray range with high angular resolution.
A particle radiation monitor (INTEGRAL Radiation Environment Monitor, IREM) measures the rates of local highly-energetic particles that can cause damage to spacecraft components. All instruments operate simultaneously. Details of each subsystem will be given in the remainder of this paper.

INTEGRAL follows a highly elliptical orbit. From launch to early 2015, one revolution around the Earth lasted 72\,hr.
In January/February 2015 the orbit was significantly modified to ensure safe disposal of the satellite in early 2029 (see Sect.~\ref{sec:Conclusions}).
The new orbit has a 64-hour duration, i.e., 3 revolutions in 8 days.
The evolution of INTEGRAL's orbit led to a peak perigee altitude of about 9550 km in Autumn 2015. Since then, the altitude has been decreasing again with another minimum of $<$2000\,km in 2020.
The real-time nature of the INTEGRAL mission requires full ground-station coverage of the operational orbit.
Ground-station coverage is currently achieved by ESA's Kiruna station, and augmented by other stations when necessary.
The satellite requirements of the orbital scenario are dictated by power consumption, thermal requirements and operational considerations. In order to guarantee sufficient power throughout the mission, the Solar aspect angle is currently constrained to $\pm$40$^\circ$.
This implies that the pointing angle of the spacecraft must be greater than 50$^\circ$ away from both the Sun and the anti-Sun.

In order to minimize systematic effects due to spatial and temporal background variations in the IBIS and SPI instruments,
a controlled and systematic spacecraft dithering maneuver is required. This maneuver consists of several off-pointings of the
spacecraft's pointing axis from the target in 2.17$^\circ$ steps. The integration time for each pointing ('science window') on this raster scan is
between 30 and 60 minutes, adjusted such that an integer number of complete dither patterns is executed.
There are three distinct observation modes: rectangular (so-called 5x5) dither, hexagonal (so-called HEX) dither, and staring (no dither).
Over the years various user-customized patterns have also been used. During all observations, the spacecraft provides stable pointings within
7.5" of the pointing direction. The only mode suitable for deep exposures is the rectangular (5x5) mode, as it is the most effective in reducing
instrumental artifacts in stacked images, spectra, and light curves.

The scientific goals of INTEGRAL are obtained using relatively high-resolution spectroscopy, combined with fine imaging and accurate positioning of celestial gamma-ray sources, allowing identification with counterparts at other wavelengths. Moreover, these characteristics can be used to distinguish extended emission from point sources and thus provide considerable power for serendipitous science: a very important feature for an observatory-class mission.
Routine INTEGRAL science operations are implemented using a pre-planned sequence of observations as a baseline. The X- and gamma-ray sky is, however, highly variable and many of the mostly unpredictable ToO events are scientifically important and so often warrant modifying the
pre-planned observing schedule.
INTEGRAL has no on-board gamma-ray burst (GRB) detection and triggering system. However, it continuously downlinks its acquired data to Earth,
allowing for constant, near-real time burst monitoring on the ground. The 'INTEGRAL Burst Alert System' (IBAS)
sends out alerts with the characteristics of GRBs detected in the FoV of the main instruments.
The SPI Anti-Coincidence (veto) Subsystem (ACS), with a time resolution of 50\,ms and time-tagging down to 1\,ms at energies above 75\,keV, and
the IBIS VETO system, with a time resolution of 8\,s at energies above 50\,keV,
provide an all-sky monitor for GRBs. For instance, SPI-ACS detects about 300 GRBs per year outside the
FoV of the instruments.
Therefore, INTEGRAL is a key participant in the Inter-Planetary Network (IPN) that combines data from several satellites to provide
accurate localisations of GRBs, crucial for follow-up observations at other wavelengths.
Apart from finding GRBs, INTEGRAL is well-suited to the search for any transient electro-magnetic counterpart to, e.g., gravitational-wave (GW) signals, ultra-high energy neutrino (UHEN) events and fast radio bursts (FRBs), owing to the large FoVs of IBIS and SPI, and their nearly omni-directional anti-coincidence shields.

INTEGRAL Announcements of Opportunities (AOs) for observing proposals are open worldwide. The provision of the Proton rocket by ROSCOSMOS led to an optimised orbit that increased the available observing time as opposed to earlier estimates. As a result, currently, 25\%\ of the Open Time is reserved for scientists from the Russian Federation. ESA provides the opportunity to propose coordinated observations, with ESA's XMM-Newton, NASA's NuSTAR and NASA's Neil Gehrels Swift Observatory missions. INTEGRAL also has a very strong Target Of Opportunity policy which enables any observers, from different astronomical domains, to propose  observations of transient events.

The INTEGRAL Science Working Team (ISWT) was formally established at the time of the selection of the INTEGRAL payload, data centre and mission scientists
(1995). In 2005, ESA decided to set up an INTEGRAL Users' Group (IUG) in parallel to the ISWT.
After taking the decision to open up all available observing time to Guest Observers from 2009 onward, the ISWT was formally dissolved in 2007,
merging with the INTEGRAL Users' Group (IUG) that acts as an interface between the scientific community and the mission management.

Many details about the INTEGRAL spacecraft, orbit, instruments, scientific aims and first results can be found in A\&A Volume 411 (2003), a special Astronomy \&\ Astrophysics issue dedicated to INTEGRAL; these papers are still valuable references.
In this paper, we present updates on the status of the spacecraft and its instruments, as it evolved during the mission, and include some aspects of ESA's Ground Segment.
The rich science which resulted from INTEGRAL's observations over more than 18 years are presented in the various review articles included in this Volume.

\section{SPectrometer on INTEGRAL - SPI}
\label{sec:spi}
\subsection{SPI main instrument}

SPI is a spectrometer telescope sensitive in the 20\,keV--8\,MeV energy range,
the hard-X-ray/soft-gamma-ray domain, with high
energy resolution at the keV level (about 3\,keV at 1.8\,MeV, i.e., $\Delta$E/E$\simeq$0.18\%).
The core of the instrument consists of 19 High Purity Germanium (HPGe) detectors, cooled to 80\,K, and surrounded by an active anti-coincidence shield (ACS).
Modest imaging capabilities are provided by a coded mask, with a design adapted to the configuration of the 19 detector elements.
The ACS identifies and suppresses prompt background that is created within the spacecraft and instrument by incident cosmic-ray particles; due to its high and omni-directional sensitivity, the ACS also functions as a detector for transient gamma-ray sources such as gamma-ray bursts (see Sect.~\ref{sec:spiacs} for a detailed description of the SPI-ACS).
Time-tagging of the SPI detector events with an internal precision of 50\,ns and an effective absolute timing precision of 50\,$\mu$s
allows for period folding analysis such as required for pulsar signal searches.
The modular design of the SPI camera also enables studies of polarization of incident gamma-rays.
The SPI instrument design and its components are described in \citet{Vedrenne2003}.
Its in-flight performance after launch is presented in \citet{Roques2003}.
A study of response and background characteristics over 13.5 years is presented in \citet{Diehl2018}.
A detailed discussion of instrument behaviour and the impact on event selection for spectra of bright sources is given in \citet{Roques2019}.
A study of the Crab Nebula emission over the mission lifetime \citep{Jourdain2020} demonstrates that the SPI instrument efficiency remains within 5\% of its initial value since launch.

Four of the 19 germanium detectors (GeDs) experienced a failure of the Ge pre-amplifier during the early mission (see Table \ref{table:GeD}). The exact cause of these anomalies remains unknown; one hypothesis is the destruction of the field-effect transistors (FETs) due to a huge
energy deposit (possibly heavy ions), or a transient in the high voltage supply. No further failures have occurred since May 2010.

In space, the constant irradiation by cosmic-ray particles (mainly neutrons and protons) degrades the detectors' crystalline lattice.
The operating temperature has a significant effect on the energy resolution of an irradiated (degraded) GeD.
The initial operating temperature was set to 90\,K after launch and in-orbit verification (see Table~\ref{table:GeD}).
Given the excellent performance of the whole cooling system,
it was decided to lower this temperature in order to reduce the rate of detector degradation. This has been done in a few steps
(see Table \ref{table:GeD}). Since October 2006, the operating temperature of the detectors is 80\,K.

\begin{table*}
	\centering
	\caption{SPI camera configuration milestones during the INTEGRAL mission}
	\begin{tabular}{rcccc}
		\hline
		\hline
		\multicolumn{1}{c}{Rev$^1$} &
		\multicolumn{1}{c}{Date} &
		\multicolumn{1}{c}{GeD} &
		\multicolumn{1}{c}{Operating} &
		\multicolumn{1}{c}{Change/Anomaly} \\
		\multicolumn{1}{c}{\#} &
		\multicolumn{1}{c}{} &
		\multicolumn{1}{c}{temperature} &
		\multicolumn{1}{c}{Voltage} &
		\multicolumn{1}{c}{} \\
		\multicolumn{1}{c}{} &
		\multicolumn{1}{c}{} &
		\multicolumn{1}{c}{[K]} &
		\multicolumn{1}{c}{[kV]} &
		\multicolumn{1}{c}{} \\
		\hline
		8    & Nov 2002 & 90 & 4.0 & initial ops configuration\\
		44   & Feb 2003 & 85 & 4.0 & temperature change\\
		140  & Dec 2003 & 85 & 4.0 & GeD \#2 failure\\
		215  & Jul 2004 & 85 & 4.0 & GeD \#17 failure\\
		455  & Jul 2006 & 82 & 4.0 & temperature change\\
		492  & Oct 2006 & 80 & 4.0 & temperature change\\
		776  & Feb 2009 & 80 & 4.0 & GeD \#5 failure\\
		930  & May 2020 & 80 & 4.0 & GeD \#1 failure\\
		982  & Oct 2010 & 80 & 3.0 & HV change\\
		1161 & Apr 2012 & 80 & 2.5 & HV change \\
		\hline
		\multicolumn{5}{c}{\footnotesize{$^1$Rev = INTEGRAL Revolution}} \\
	\end{tabular}
	\label{table:GeD}
\end{table*}

\begin{figure*}[ht!]
    \includegraphics[width=\linewidth]{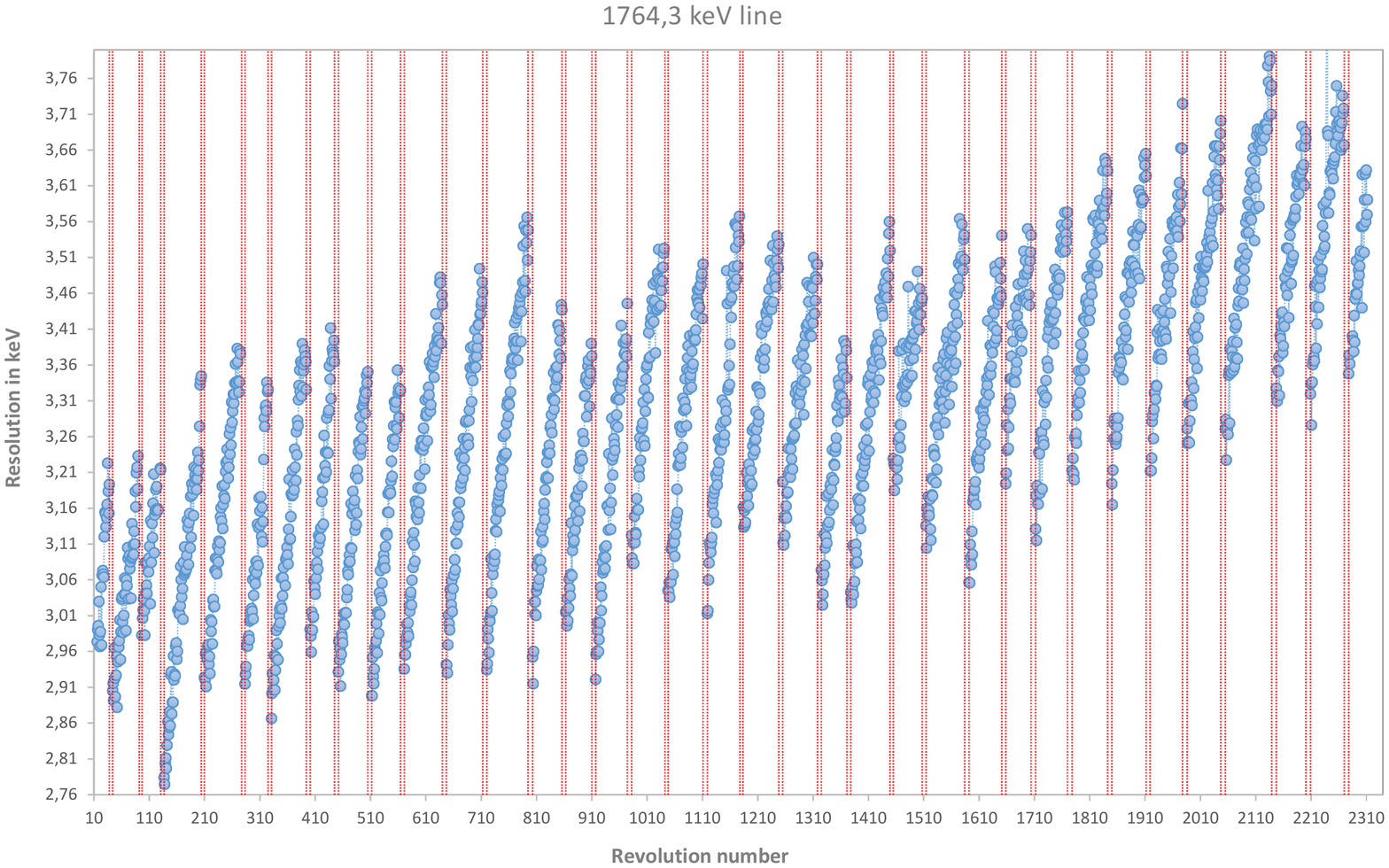}
	\caption{Evolution of SPI's energy resolution during the mission for the $^{205}$Bi background line at 1764.3\,keV, up to December 2020 (Revolution 2312). Red vertical lines indicate annealing periods.}
	\label{annealing1}
\end{figure*}

The irradiation by cosmic-ray particles results in a continuous worsening of the detector energy resolution, by typically 10\%\ every
6 months, if no corrective action is performed (see Fig.~\ref{annealing1}).
SPI has been designed to allow annealing of the GeDs. This operation consists of heating up the detectors to 103$^\circ$C.
This restores the quality of the crystal lattice, so that charge collection and the original energy resolution is recovered.
Annealing is a critical operation which puts a tremendous thermal stress on the detector with a thermal excursion of almost 300 degrees.
Since launch, SPI has been continuously monitored, tuned and improved (such as GeD operating temperature, annealing interval,
annealing duration) in order to keep its scientific performance at the highest possible level.
The annealings are performed roughly every six months, each with an optimized duration of about 200\,hr;
this is far beyond original mission planning and any qualification limit.
Fig.~\ref{annealing1} displays the evolution of the energy resolution for the $^{205}$Bi background line at 1764.3\,keV versus
INTEGRAL Revolution number. This graph shows that the annealing strategy succeeded in limiting
the deterioration of the effective mean energy resolution between 2002 and 2019 to $\lesssim$15\%.

The cryogenic system is a key SPI component and consists of 2 pairs of Stirling coolers working in parallel, to achieve an operational temperature of 80\,K.
Between annealings, the cooling efficiency degrades, possibly due to contamination of the insulation by water.
During the annealings, the cold plate is warmed up to 103$^\circ$C.
As a consequence, the surfaces are ``cleaned'' and the efficiency of the cryogenic system is restored to its initial value at the beginning of the next cooling period.
However, the cryogenic system works nominally with unchanged performance and the cold plate temperature stability is controlled within 1$^\circ$C.

The SPI Digital Processing Electronics (DPE) unit manages the SPI sub-assemblies and
the interface with the spacecraft for the commands and the telemetry. Since launch, a few bugs have been corrected and improvements to simplify operations have been implemented in the on-board software. Due to the limits on INTEGRAL`s telemetry,
significant efforts have been undertaken to optimize SPI telemetry usage, reducing the SPI global telemetry rate by more than 30\%. The counting rate increase due to the background increasing with
the Solar modulation was compensated in steps with different compression algorithms, including the compression of all type of events, implemented at the end of 2009.

Soon after launch, it became clear that ``spurious'' events were produced by the electronics. The most prominent fraction of these (electronic noise) events are located around 1.4\,MeV and have appearances similar to instrumental lines, but lower energies are also affected \citep{Jourdain2009}.
As a consequence of the electronic saturation, an effect similar to pile-up shifts low energy photon events toward higher apparent energy.
This is particularly relevant for high-intensity sources exceeding a large fraction of the flux of the Crab, for which the fraction of displaced photon events from low to high energy can significantly affect the observed spectral shape above $\sim$400\,keV.
Such pile-up issues can be solved through proper event selection at the data processing level \citep{Roques2016, Roques2019}, available to general users as part of the Off-line Science Analysis (OSA) software\footnote{Since OSA release 11. See \url{http://isdc.unige.ch/integral/download/osa/doc/11.1/osa_um_spi/node50.html}.} provided by the
INTEGRAL Science Data Centre (ISDC; see Section~\ref{sec:isdc}) or via the SPI Data Analysis Interface (SPIDAI) website\footnote{~\url{https://sigma-2.cesr.fr/integral/spidai}}.

The instrument response can be separated into a geometrical part (the imaging response function, IRF, which implements the coded-mask properties and uses the INTEGRAL Mass Model (\citep{Ferguson2003}))
and the spectral response of the GeDs. The latter can be implemented as an energy redistribution model via the redistribution matrix file, RMF, which reproduces photon interactions in the SPI detector plane and
surrounding material.
The IRF and RMF response matrices have been built from simulations, and compared with ground calibrations \citep{Sturner2003, Attie2003}.
They are applied without in-flight adjustments, but have been updated/re-built on six occasions,
four of them corresponding to the time of detector failures (see Tab.~\ref{table:GeD}).
A first revision was also required at low energies, to better reproduce the observed decrease of the efficiency below 60\,keV \citep{Sturner2003}.
Another update in May 2005 accounted for obscuration effects of the JEM-X mask support structure which appear when a source is located at specific angles (i.e., around 12$^{\circ}$ or more off-axis towards the JEM-X/IBIS side, see Fig.~\ref{fig:integral}).

INTEGRAL regularly observes the Crab for calibration purposes (currently a short, 12.5-hour, observation about every 4 revolutions, and a longer observation, 2 revolutions long, twice per year). This source is considered to be a standard candle
in SPI`s energy range, even if 5--7\%\ variations per year have been observed around a stable long trend \citep{Wilson-Hodge2011}.
SPI offers the opportunity to study this source using only ground-calibration knowledge:
a long-term study incorporating the first 6 years demonstrated the stability of the Crab (Nebula) emission
in terms of spectral shape and intensity, together with the stability of the instrument response, within $\sim$5\%\ uncertainty \citep{Jourdain2009}.
A more recent study, using results obtained from the 2016--2018 Crab observations were compared to those from 2003, and
led to an improvement of the determination of the Crab spectral shape \citep{Roques2019}, modelled with a Band model rather than a
broken power law. \citet{Jourdain2020} showed that the source flux variability appears to be contained within less than
$\pm5$\% around a $\sim$20\,yr mean value in a broad band covering the 20\,keV--400\,keV energy domain and that
the instrument efficiency remained within the same range.

The dominating instrumental background underlying the source signal, and also the instrument's imaging response to the gamma-ray sky, require careful
implementation of the spectral and coded-mask response functions and the instrumental background model during the scientific data analysis.
A direct deconvolution of the SPI data is challenging, because the response matrix is singular and inversion is ill-defined.
Background signal is modelled together with the source
in the same data space of detectors and their counts per energy bin.
The response to background is recorded by the active detectors, since background from
all directions imprints its pattern on the detector.
This pattern remains fairly constant over periods of days \citep{Diehl2018}, despite changing its level continuously.
This offers the possibility for broad-band spectral bins to use observations of regions without sources to
produce a background model (\emph{flat fields}) that can be scaled to match the background level in each
science window.
Exploiting only the signal modulation from the coded mask and the dithering strategy, over a sufficiently long observation, it is possible to disentangle the signal of the source from the overwhelming background.
Beyond such flat-fielding, the SPI teams have developed other methods and tools to extract scientific products, allowing more elaborate background modeling.
As an example, it is possible to fit the background detector pattern from data at high spectral resolution, and it is possible to
use scaling from radiation monitor detectors to account for short-term background variations.
These methods have been used in various works \citep[see][and references therein]{Diehl2020}

\begin{figure*}[ht!]
	\centering
	\includegraphics[width=0.80\linewidth]{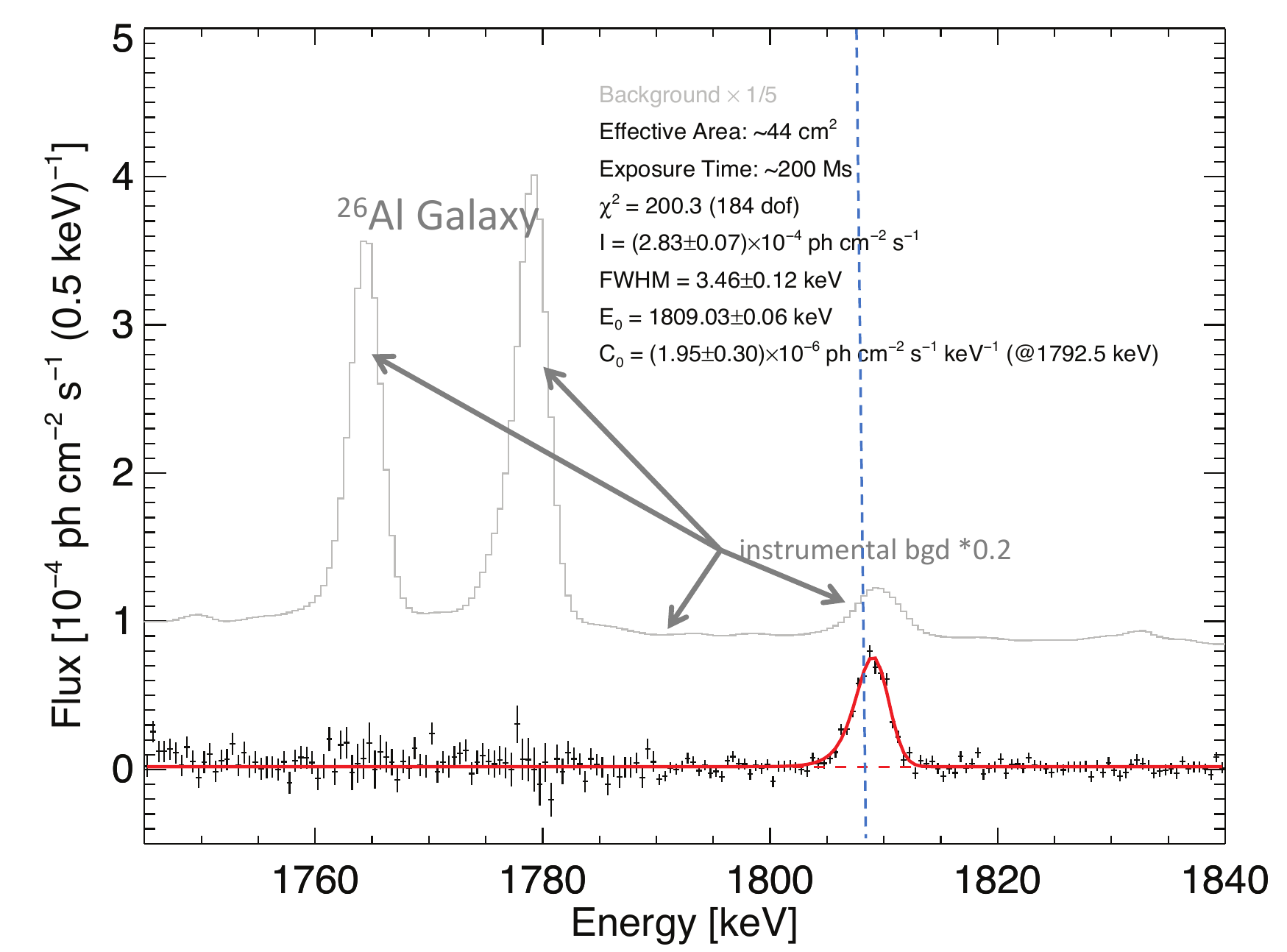}
	\caption{Astrophysical result from SPI event data analysis for diffuse lines. Background suppression example for the case of radioactive decay of $^{26}$Al \citep{Siegert2017}. The raw count spectrum shown in grey illustrates the background lines; the result of a model-fitting analysis for the $^{26}$Al emission from the galaxy is shown as fitted coefficients per energy bin, including uncertainties. The instrumental lines near 1764 and 1778\,keV are properly removed, with minor systematics fluctuations remaining.
}
	\label{fig:26Al_bgdsuppression}
\end{figure*}

\begin{figure*}[ht!]
	\centering
	  \includegraphics[width=0.8\linewidth]{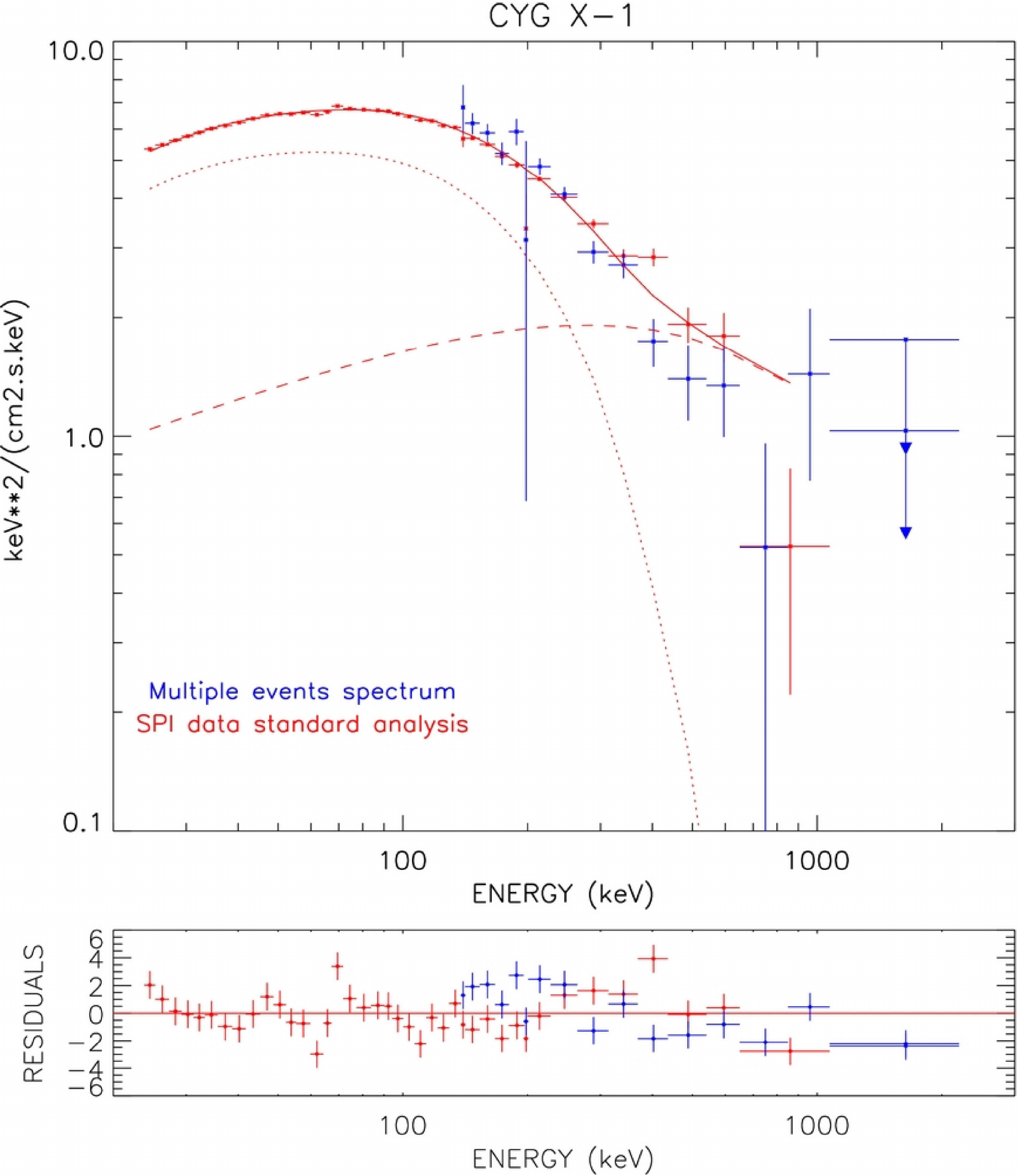}
	\caption{Astrophysical result from SPI event data analysis for point source emission. Cygnus X-1 stacked SPI spectra obtained using standard analysis (single events; 22\,keV--2\,MeV, red points) and multiple events (130\,keV--8\,MeV, blue points). The solid lines represent the best-fit model composed of a thermal Comptonization (\texttt{reflec*CompTT}, dotted curve) plus a fixed cut-off power-law high-energy component (dashed line). From \citet{Jourdain2012}.
}
	\label{fig:SPI-Cyg}
\end{figure*}

Using these methods, it is possible to obtain the intrinsic signal both for diffuse emission in narrow lines and the continuum spectrum of point sources
(see Figs.~\ref{fig:26Al_bgdsuppression} and \ref{fig:SPI-Cyg} for two examples).
For narrow astrophysical gamma-ray lines, a sensitivity of $\sim$10$^{-5}$\,ph\,cm$^{-2}$\,s$^{-1}$ has been achieved (for 10$^6$\,s observing time).

\subsection{SPI as a polarimeter}

Thanks to its independent crystals, SPI can be used as a polarimeter. Above $\sim$130\,keV,
a significant number of photons undergo Compton scattering in one detector, with the diffused photons being absorbed
in the next one (so-called `Multiple events' or ME). Since during a Compton interaction the photon is preferentially
diffused (in azimuth) in a plane perpendicular to the polarization vector (= Electric vector), the spatial distribution of the
ME contains the polarization information of the incident flux.
The effective area of SPI corresponding to ME varied from 10 to 50\,cm$^{2}$ between 150\,keV and 1\,MeV at the beginning of the mission \citep{Attie2003}.
The failures of 4 detectors during the mission resulted in a progressive decrease in the ME effective area down to $\sim$52\% of the original area.

SPI`s first polarimetric studies were devoted to a gamma-ray burst (GRB), GRB041219A \citep{McGlynn2007}, then to the Crab off-pulse emission \citep{Dean2008}.
Since then, the analysis tools have been substantially improved.
A detailed description of the method is given in \citet{Chauvin2013}, in which they applied their method to the total Crab (pulsar + nebula) emission. A polarization fraction of $\sim$30\%\ for an angle of 117$^\circ$,
aligned with the spin axis of the pulsar was found, in agreement with the SPI and IBIS results from \citet{Dean2008} and \citet{Forot2008}, respectively.

\section{Imager on Board the INTEGRAL Satellite - IBIS}
\label{sec:ibis}
\subsection{IBIS main instrument}

IBIS is the high angular resolution gamma-ray imager on-board INTEGRAL (\citep{ubertini2003}).
It operates in the range 15\,keV--10\,MeV with $<$120\,$\mu$s time resolution,
and based on two independent solid state pixellized detector arrays optimized for low (15--1000\,keV; ISGRI: 'Integral Soft Gamma-Ray Imager'; \citealt{Lebrun2003})
and high (0.175--10.0\,MeV; PICsIT: `PIxelated CsI Telescope'; \citealt{Labanti2003}) energies, shielded by an active Veto System cage protecting 5 sides of the detection array \citep{Quadrini2003}.
This shield is essential to minimise the background induced by high-energy particles along INTEGRAL's highly eccentric orbit.
The imaging capability is obtained with the use of a tungsten Coded Aperture Mask, projecting the sky images on the IBIS detection planes, ISGRI and PISCsIT.
IBIS is characterized by good angular resolution (12\,arcmin) and unprecedented point-source location-accuracy (PSLA) of 1--3\,arcmin for moderately
strong sources (typical fluxes of a few mCrab), reaching a few arcseconds for bright sources. In addition, the IBIS optical system features a
collimation system, connecting the detection plane to the mask, in order to shield the two detection layers, ISGRI and PICsIT, against low energy
particles from cosmic rays, diffuse galactic and extra-galactic X-rays and soft gamma-rays, moderate Solar flares and strong contaminating sources
out of the Field-of-View (FoV) of the telescope. This shielding effect is obtained with the use of a hopper, placed on top of the detector system.
This hopper is then connected to the mask with a lead tube with a variable thickness extending up to the mask.
A lead door complements the shield in the direction opposite to the SPI. This results in a 9$^{\circ}$$\times$9$^{\circ}$ fully-coded FoV, surrounded
by a 30$^{\circ}$$\times$30$^{\circ}$ partially-coded FoV, with decreasing exposed area from 100\% to 0\%.
The whole system, becoming increasingly transparent for energies above 200\,keV, was designed to reduce the diffuse sky component of the
instrument background (see \citealt{Savchenko2017}, and references therein, for more details).
Finally, the reconstruction of the sky images is
obtained via cross-correlation between the detector images and the coded mask array function as described in \citep{Goldwurm2003}, where  
the standard IBIS data analysis implemented in the OSA software is reported.
IBIS is also sensitive to off-axis high energy photons via the ``Compton Mode'', with a moderate angular resolution capability (\citealt{forot2007}).

ISGRI is usually operated in photon-by-photon mode, while, due to the anticipated high energy excess counting rate and limited telemetry bandwidth, PICsIT is routinely operated in ``Imaging Mode'', integrating sky images with a duration from 0.5 to 2\,h. To overcome this limitation, and improve the high-energy
sensitivity (E$>$200\,keV) to transient events, a high-time resolution mode, namely the "Spectral Timing Mode", is operational for PICsIT,
transmitting the number of photons detected over 8 energy channels with integration time normally set to 7.8\,ms \citep{Labanti2003}.

ISGRI consists of 128x128 pixels of semiconductor detectors. Photons with energy above 100\,keV can
interact in the full depth range of each pixel. When the interaction happens deep in the detector, it
takes longer for the charge carriers to reach the electrodes and the registered pulse peaks at
a later time: the delay is dominated by the transport of the holes. Due to charge-trapping in the
detector medium and the increase of the ballistic deficit when the pulse duration increases,
deeper interactions are also registered with smaller pulse heights. If not corrected, this effect leads
to a substantial decrease of the energy resolution. To solve this problem, ISGRI pioneered using a bi-parametric approach:
measuring both pulse height and rise time for each energy deposition event and relying on on-ground software to perform the corrections \citep{Lebrun2003}.

In former versions of the Off-line Science Analysis (OSA) software, the time evolution of detectors' gain and offset (split in two  energy regions)
was modelled using a phenomenological description \citep{Caballero12}. The bi-parametric spectra were corrected to make the positions of the background lines
at 60\,keV and 511\,keV compatible with those at the beginning of the mission. This correction did not take into account the evolution of the energy
resolution and low-energy threshold. It was not possible to predict changes in the instrument detection efficiency, so the spectral response normalization was adjusted to make the Crab spectrum consistent with the INTEGRAL/SPI measurements.

In the last few years, a more complete time-dependent physical model for the charge loss in the detector has been developed (Lebrun et al.\ 2021, in preparation).
The charge loss due to finite carrier mobility, ballistic deficit and resolution of the measuring electronics was adjusted using
ground calibration data and in-flight line properties. This model predicts the dependency of the line position on the rise time
throughout the mission with good accuracy. The two-dimensional model line profiles are fitted to the bi-parametric data for each INTEGRAL
revolution. This response model, describing the distribution of final measured photon energies for each energy of the incident photon,
enables one to construct energy corrections and response matrices which now take into account these aspects of the ISGRI response time variability.
This is integrated in the latest OSA version (OSA11; release date: October 2018), and the response matrices are being updated for the whole mission.

The major effects of the OSA11 improvement are the possibility to measure spectra with a non-trivial shape (e.g., with a cyclotron absorption line), to reconstruct the absolute flux of hard X-ray sources, and to have source average spectra which are now consistent with other
hard X-ray instrument measurements, enabling better cross-calibrations.

\subsection{IBIS as a polarimeter and Compton telescope}

In all gamma-ray (Compton) polarimeters, the dependency of the polarization on the differential scattering cross section is used to constrain the degree
and angle of linear polarization of the incident radiation. Indeed, linearly polarized photons scatter preferentially perpendicular to the incident
polarization vector. Hence by examining the azimuthal distribution of scattered photons among different detectors of the same instrument, one can in principle derive the degree and angle of linear polarization of the incident gamma-ray photons.
Thanks to its two position-sensitive detectors, ISGRI and PICsIT, IBIS can also be used as a Compton polarimeter to study many compact objects.
The procedure to measure the polarization is described in \citet{Forot2008}. It allows control of systematic effects and successfully
detects polarized signals from bright sources, such as the Crab nebula, Cygnus X-1 and many GRBs.

Compton telescopes are generally used to image the sky between a few hundred of keV to several MeV, thanks to the Compton kinematics information.
The main advantage of these telescopes is their inherent low background; however, their energy-dependant spatial resolution and background subtraction
methods introduce complications. One way to overcome these problems is to use a coded aperture mask together with the Compton reconstruction.
Indeed, coded mask imaging allows an automatic subtraction of background events, keeping an energy-independent angular resolution.

IBIS is routinely used as a Compton coded-mask telescope. True Compton scattering occurs when two events are detected in ISGRI and in PICsIT.
Among these events, only a few are true Compton events coming from a source. A large fraction are due to background Compton events, while another part is due
to spurious events. It may happen indeed that an event recorded by ISGRI is mistakenly associated with an independent event detected by PiCsIT, when
both events are detected within a time window small enough for the IBIS on-board software to consider them simultaneous.
Some of the spurious events can be source photons passing through the mask, and thus are not suppressed during the coded-mask image processing.
In the IBIS/Compton data analysis software, they are quantified and subtracted from the final sky image with high accuracy to avoid a wrong source-flux
computation.

\section{INTEGRAL as an 'All-sky gamma-ray monitor'}
\label{sec:moitor}

Starting in 2016, the scientific requirements and capabilities of INTEGRAL have been revisited.
The target was to assess and verify its expected performance for counterpart detection in coincidence with gravitational wave (GW) prompt emission,
and possibly, to detect gamma-ray afterglows (see, e.g., \citealt{vandenheuvel2017}).
INTEGRAL benefits from long, uninterrupted observations (about 170\,ks per satellite orbit of 2.7 days), with an instrument outage of only about 6\,h during perigee passage, giving an operational duty cycle of $\sim$85\% with negligible occultation by the Earth.

The field of view of the main instruments covers with optimal sensitivity a relatively small region of the sky for
serendipitous search. However, the anti-coincidence shields of SPI and IBIS are effective detectors of gamma-ray flashes, besides
shielding the detectors from cosmic rays.

\subsection{SPI Anti-Coincidence System}
\label{sec:spiacs}
The germanium detectors (GeDs) of SPI are surrounded by an active anti-coincidence shield, SPI-ACS, forming a collimator tube and a bottom shield.
SPI-ACS \citep{Vonkienlin2003} is made of 91 BGO (bismuth germanate, Bi$_4$Ge$_3$O$_{12}$)
scintillator crystals.
The SPI-ACS is very efficient in recording events due to charged particles (cosmic rays) that induce background events in the SPI detectors. Also, at low energies, the scintillator is thick enough to prevent gamma-ray photons from reaching the GeDs from directions other than the instrument's field-of-view, as defined by the mask.
Data selections to exclude high ACS count rates efficiently suppress background from solar flares or particle events in the Earth magnetosphere, as well as at times when INTEGRAL enters and exits the radiation belts during its orbit.

\begin{figure*}[ht!]
	\centering
	\includegraphics[width=0.9\textwidth]{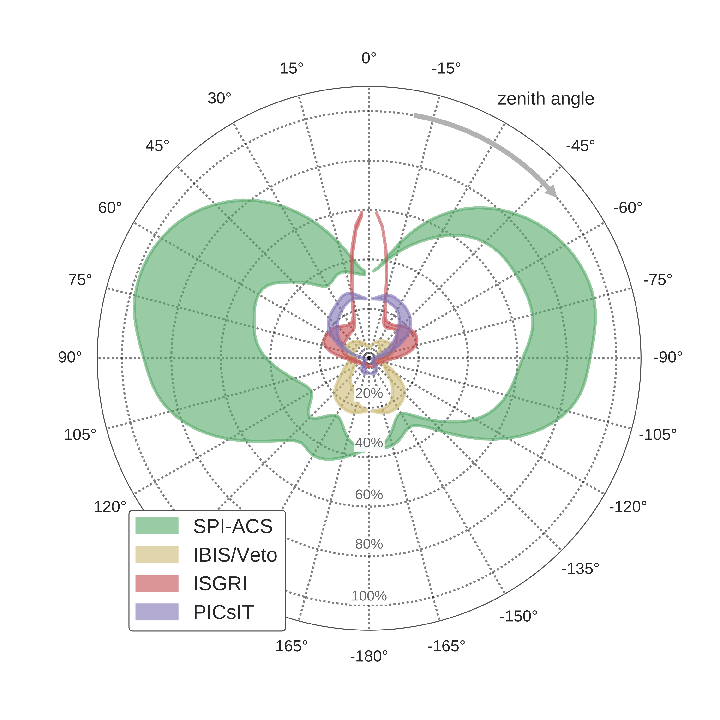}
	\caption{The normalized all-sky sensitivity of the SPI-ACS to a typical short gamma-ray burst lasting 1\,s, as a function of zenith angle and compared to other transient-source monitoring instruments on INTEGRAL. The axis between IBIS and SPI is at an azimuth of 90$^{\circ}$. The shaded areas represent the typical azimuthal range of sensitivities  (from \citealt{Savchenko2017}).}
	\label{fig:ACSresponse}
\end{figure*}

In addition to its main function of providing a veto signal for charged
particles irradiating the SPI instrument, the ACS also provides the
count rate of all impinging particles and high-energy photons.
It can thus be used as a nearly omni-directional detector of transient events,
with an effective area reaching 0.7\,m$^2$ at energies above 75\,keV and a
time resolution of 50\,ms \citep{Savchenko2017}.
The directional dependency of effective area
for the SPI-ACS, as shown in Fig.~\ref{fig:ACSresponse}, is strongly affected by the opacity of materials which are
part of the INTEGRAL satellite structure and other instruments \citep{Savchenko2017}. For usage as a gamma-ray detector this opacity was estimated
through a number of Monte Carlo particle transport simulations, performed
using both the SPI detector mass model and the INTEGRAL satellite
Mass Model (TIMM; \citep{Ferguson2003}). Both models predicted similar count rates for the
known impulsive events (gamma-ray bursts, soft gamma-ray repeater flares) that were observed by SPI-ACS,
within an accuracy of 20\%.

\subsection{IBIS as an all-sky monitor}
\label{sec:ibisveto}

 The active IBIS shielding covers the bottom and lateral sides of the IBIS detectors with 16 Veto Detector Modules (VDMs), in four different shapes for shield leakage minimisation \citep{Quadrini2003}. Each module comprises two BGO scintillation crystals optically glued along their long edge. The composite crystal is viewed by two PMTs with embedded Front End Electronics (FEE) and high voltage (HV) divider.  In each module a High Voltage Power Supply (HVPS) is distributed to the two PMTs whose signals are summed after adjustment. The sum is delivered to the Veto Electronics Box (VEB) where all signals are discriminated, converted into strobes with adjustable length and delay, and distributed to the two detector layers, ISGRI and PICsIT. Optimization of the system delivers an average dead-time of about 15\% for ISGRI and 4\% for PICsIT. Data are also collected in the so-called housekeeping (HK) data, which provide
 a public, continuous data stream with time resolution of 8\,s, which can also be exploited to detect gamma-ray bursts, especially those
 in the region of the sky opposite to the INTEGRAL field of view where SPI-ACS is less sensitive.

IBIS has a non-negligible off-axis response, which can also be exploited to detect serendipitous transients (see \citealt{Marcinkowski2006}, \citealt{Marcinkowski2007}, \citealt{savchenko17}, \citealt{Chelovekov2019}).
The following IBIS main features are key for the search of GW counterparts and the detection of short gamma-ray bursts (GRBs): access to an energy range
22\,keV--10\,MeV with unprecedented sensitivity\footnote{Note that the lower energy range increased from 15\,keV in the beginning of the mission to 22\,keV in recent times, due to detector in-flight aging.}; near all-sky coverage in the range 50\,keV--2.5\,MeV; high-time resolution: $<$120\,$\mu$s single photon
absolute arrival time reconstruction, 7.8\,ms PICsIT spectral timing mode; and polarimetry capability.
If a transient appears in the wide FoV of IBIS ($\sim$100~deg$^2$, fully coded -- 1000~deg$^2$, to zero response), the
point-source location accuracy (PSLA) can span from 5 arcmin to a few arcsec, depending on the source strength.
This is an exceptional asset, exploited to detect faint GRBs \citep{Gotz2019} and other possible transient signals \citep{Ferrigno2021}.
Last but not least, IBIS/Compton data may be used to make Compton images out of the coded-mask FoV. This is possible only at high energies, above 300\,keV,
when the IBIS shielding becomes transparent. This Compton image process effectively extends the IBIS FoV.

\subsection{Combining all instruments}

\begin{figure*}[ht!]
  \centering
  \includegraphics[width=\linewidth]{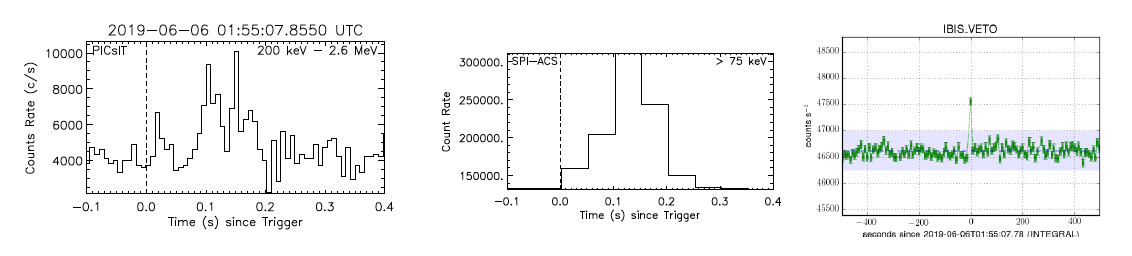}
  \caption{The short GRB190606A detected by the different ``all-sky monitor'' instruments aboard INTEGRAL.
As can be seen, PICsIT (left) provides the best time resolution over 8 energy channels (200\,keV--2.6\,MeV), SPI-ACS (middle) provides the best
sensitivity with its large collecting area with 50\,ms time resolution integrating photons with energies above 70\,keV \cite{Vonkienlin2003},
and IBIS-Veto provides good sensitivity from 50\,keV up to 2.5\,MeV with 8\,s integration time (\cite{Quadrini2003}).}
  \label{fig:ibis2}
\end{figure*}

In order to optimize counterpart searches, an extensive simulation was performed to evaluate the IBIS and SPI sensitivity to transient events
outside the FoV, with particular attention to the absorption effects of the instruments' active and passive shields and to the obscuration due to the
spacecraft materials (for details, see \citealt{savchenko17}, and references therein).
The SPI-ACS provides the best sensitivity to short GRBs, giving integrated count rates in 50\,ms time bins, while
the IBIS-Veto is more sensitive to longer GRBs, with a minimum integration time of 8\,s.
PICsIT has a better time resolution than SPI-ACS (7.8\,ms versus 50\,ms), with the capability to provide spectral information in the
range 200\,keV--2.6\,MeV in 8 energy channels.
The instrumental set-ups nicely complement each other, therefore maximising the chance to detect, for example, GRBs as counterparts to GW events, as in the case of
GW170817/GRB170817A (\citealt{savchenko17b}), and possibly 
GW190425/GRB190425A (\citealt{Pozanenko2020}). As an example,  Fig.~\ref{fig:ibis2} shows the light curves obtained from GRB190606A as detected by SPI-ACS,
IBIS-Veto and PICsIT are shown. The capability of PICsIT to produce high time-resolution spectra is exemplified in Fig.~\ref{fig:ibis2} (\citealt{rodi2018}).

\section{Joint European Monitor for X-Rays - JEM-X}
\label{sec:jemx}

JEM-X consists of two identical coded-aperture mask telescopes (JEM-X1 and JEM-X2), co-aligned with the other instruments on INTEGRAL.
The photon detection system consists of high-pressure imaging Micro-strip Gas Chambers. An exhaustive description of the instrument can be found in
\citet{Lund2003}.
Since 10 October 2010 both JEM-X units are operating together.
The micro-strip anodes with low/noisy or no activity are taken into account in the OSA analysis software (up to January 2021; JEM-X1: 63 bad anodes, JEM-X2: 64 bad anodes).

The last 10 years of JEM-X operations have seen some interesting developments in the two instruments' behaviour.
Mostly significantly, the Cadmium ($^{109}$Cd) calibration sources (radiating at 22\,keV) on both units have become so weak that they can no longer be used.
Therefore, since March 2015 all energy calibrations have been based on the Fe sources (5.9\,keV) on JEM-X1,
which track gain variations due to temperature and X-ray flux.
The gain calibration process involves an iterative series of corrections until the instruments' Xe background lines lie with 2\%\ of the ideal
29.6\,keV value. 
Despite aging of the JEM-X instruments, and an increase in the temperature sensitivity of their gain, it is still possible to determine the average instrumental gain of both instruments with a scatter of 2--3\%\ from the ideal for all revolutions. For revolutions where there is no strong X-ray source in the FOV (i.e.low count-rate), the average gain can be determined to within 1--2\%\ of ideal.\footnote{For the specifics of each revolution see: \url{http://spacecenter.dk/~oxborrow/sdast/GAINresults.html}.}
Therefore, there has been no practical degradation of the gain
determination, despite the loss of the Cd sources. This indicates that with experience and understanding of the variables that affect the JEM-X gain,
it will be possible in the future to continue without the Fe calibration sources, when these also become too weak to use.

\begin{figure*}[ht!]
  \centering
  \includegraphics[width=0.8\linewidth]{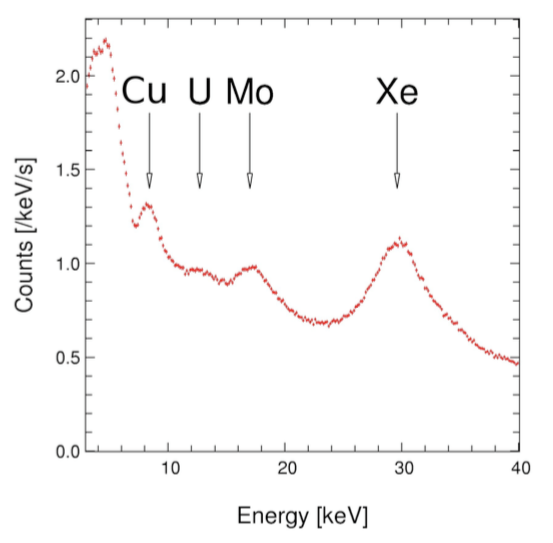}
  \caption{Full JEM-X2 detector spectrum of empty field observations with diffuse and instrumental background. The most significant fluorescent lines are indicated (the Uranium is from contamination in the Be-window).}
  \label{fig:jemX2}
\end{figure*}

The energy resolution of the JEM-X instruments degraded steadily for the first 10 years of operation, while the temperature sensitivity grew
steadily stronger. The width of the Xe background line doubled during this time. However, both these trends levelled off after about 10 years of
operation, and no significant increase in either line widths or temperature dependence has been seen since.
The major contribution to the increased line widths is due to spatial variations in the instrument gain across the micro-strip plates, and
the prevalence of cosmic-ray-induced gain glitches which are very localized. However, the stabilization of spatial gain variations and
temperature dependence, means that the JEM-X instruments can continue producing data of the same quality as 7 years ago, for the foreseeable future.

Issues with the JEM-X Crab calibration have been noted - especially the low-energy sensitivity of JEM-X2. This problem is solely due to the lower
high-voltage (HV) setting used by JEM-X2. With time, the overall gain of both instruments has dropped, and for JEM-X2 this led to low-energy
X-ray photon signals being filtered out of the data stream by the instrument's low-level discriminator. 
In May 2018 (Revolution 1948), the JEM-X2 HV was set up to the same value as for JEM-X1, to reflect that both instruments had aged a similar amount with respect to gain variations and temperature dependence. This improved the JEM-X2 performance since fewer low-energy photons are now lost in the low-level discriminator, which improved the subsequent Crab calibrations to determine instrument sensitivity, especially at the lower end of the spectral range. This improvement can be seen from May 2018 and onward. The low electronic efficiency of
both units at low energies makes it very hard to determine whether the low-level discriminator values can be lowered to benefit the lower end of the
spectral range, without swamping the telemetry stream with electronic noise.

An imaging technique (pixel-illumination-function imaging, or PIF-imaging) has been developed, which improves the visibility of weak sources close to strong sources in the JEM-X images and
mosaics. In images produced in earlier periods of the mission, the image noise increased significantly around strong sources like, for example, GRS\,1915+105
or when imaging the Galactic bulge. The PIF-imaging technique suppresses this effect. The method is based on a weighted back-projection of the detector pixels,
where the weights depend on the illumination of each pixel by strong sources.

The conclusion is that with careful monitoring of the instruments' behaviour, JEM-X can continue to deliver science results as good as those for the last 7 years of the mission, for several years to come.

\section{Optical Monitoring Camera - OMC}
\label{sec:omc}
The Optical Monitoring Camera (OMC) was designed to observe the optical
emission from the prime targets of the gamma-ray instruments aboard
INTEGRAL \citep{mashesse2003}. This capability has provided essential diagnostic information on
the nature and the physics of the sources over a broad wavelength range.
Its main scientific objectives are: (1) to monitor the optical emission
from the sources observed by the gamma- and X-ray instruments, measuring
the time and intensity structure of the optical emission for comparison
with variability at high energies, and (2) to provide the brightness and
position of the optical counterpart of any gamma- or X-ray transient taking
place within its field of view.  The OMC is based on refractive optics
with an aperture of 50\,mm focused onto a large format CCD (1024$\times$2048 pixels)
working in frame transfer mode (1024$\times$1024 pixels
imaging area). The optics includes a Johnson V filter to allow for easier
combination with photometric data taken on ground or by other space instruments.
With a FOV of 5$^{\circ}\times$5$^{\circ}$ it is possible
to monitor sources down to magnitude V=17. Typical observations
perform a sequence of different integration times, 10\,s, 50\,s and 200\,s to increase the dynamic range, allowing for photometric uncertainties below 0.1 magnitude for objects with V$\le$16. In this way, variability patterns ranging from
minutes or hours, up to months and years have been monitored. For bright
sources, fast optical monitoring at intervals down to 3\,s are possible.
The OMC is calibrated in orbit every few weeks, including a flat-field (FF) calibration using a set of on-board LED lamps.

\subsection{OMC status and evolution}

The OMC has survived satisfactorily during the long time since the launch of the mission.
The overall sensitivity has remained very stable, with a decrease below 0.02 mag of the photometric calibration zero point.
This indicates that both the optics and the CCD itself have survived very well the radiation accumulated over the years of operation.
Nevertheless, some ageing effects have appeared over time.

\begin{figure*}[ht!]
  \centering
  \includegraphics[width=0.45\linewidth]{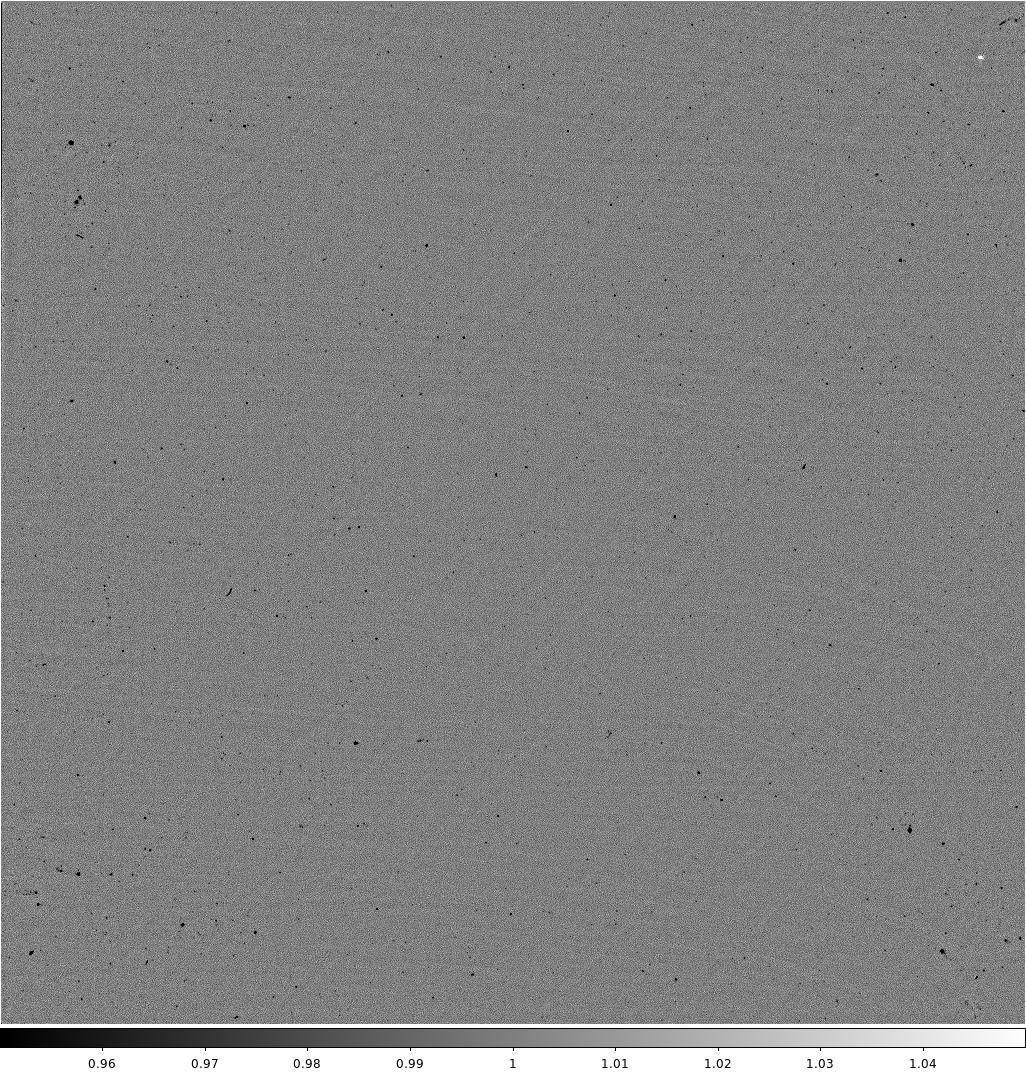}
  \includegraphics[width=0.45\linewidth]{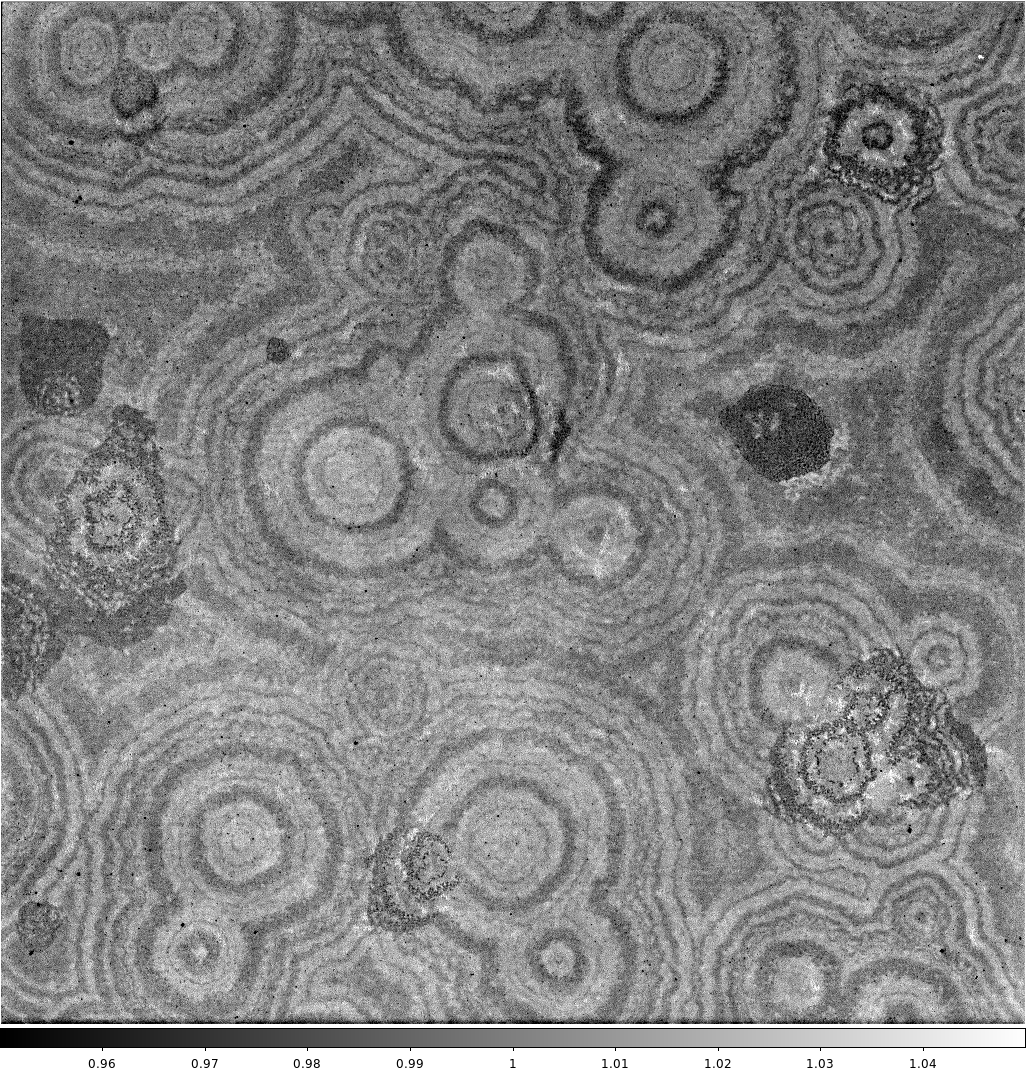}
   \caption{Evolution of the OMC Flat Field matrix from the start of the mission in 2002 (left) to mid 2019 (right). Note that the scale has been set to $\pm5$\%.}
   \label{OMC1}
\end{figure*}

The most important effect appeared a few months after launch. The FF matrix started to show rapid variations, with an amplitude of at most a few \%,
first alternating increases and decreases of the overall sensitivity over a few weeks. Then a specific pattern of more/less sensitive rings developed,
which stabilized after 5 years; it has remained stable since then. The origin of the variations has been attributed to the alteration of the
anti-reflecting coating on the CCD by the deposition of a molecular layer of contaminants, based on the fact that the CCD surface has the
lowest temperature within the focal plane volume. In Fig.~\ref{OMC1}, a comparison of the FF matrices at the beginning of the mission and
after its stabilization is shown. To achieve a proper calibration of the OMC a much more sophisticated procedure than initially planned was developed,
combining the use of LED images with the zodiacal light of specially defined sky background images. This procedure came into effect after 13 March 2004. Since then, some updates were implemented to accommodate the evolution of the instrument. The most recent was the implementation of
a narrow $3 \times 3$ dithering pattern in the OMC calibration observations for the acquisition of the sky background images,
which was included on 9 November 2016 (Revolution 1746).

The overall dark current has increased slowly, but continuously, reaching now on average 1 digital count in long exposures of 200\,s.
This is still low enough compared to the dynamic range of the instrument, up to 4096 digital counts. All CCD columns remain operational, but the number of hot (or flickering) pixels has increased to around 4000, still below 0.5\% of the $10^6$ available pixels.

\begin{figure*}[ht!]
  \centering
  \includegraphics[width=0.7\linewidth]{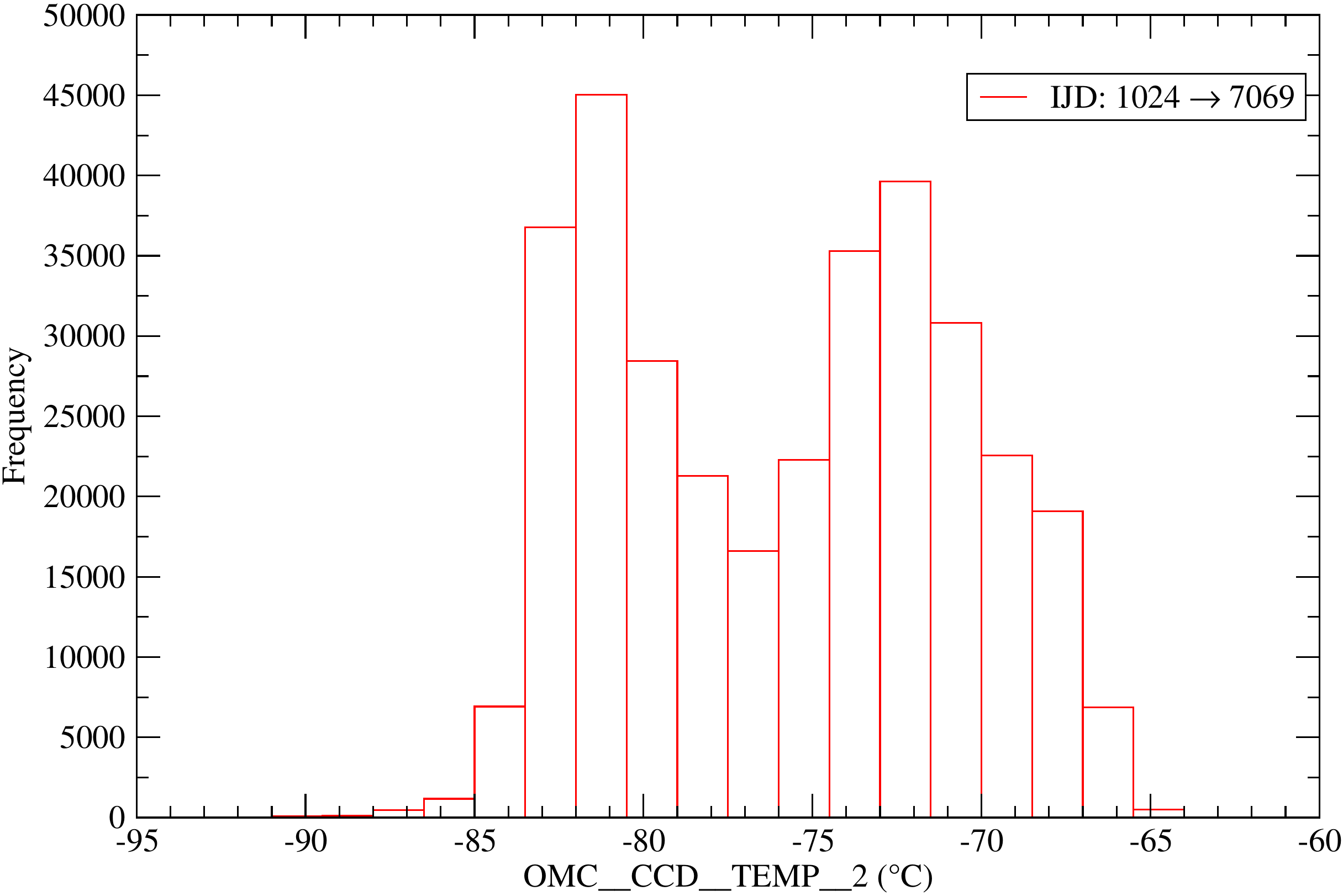}
   \caption{Histogram of OMC CCD temperatures from the beginning of the mission to mid 2019, with 2 clear peaks corresponding to the most common hot and cold attitudes of the spacecraft.}
   \label{OMC1-2}
\end{figure*}

The CCD is cooled passively via a small radiator, coupled to it by a flexible thermal strap which accommodates a thermistor.
The temperatures measured by this thermistor have remained in the range $-87$\,$^{\circ}$C to $-66$\,$^{\circ}$C since launch, with 2 peaks in the distribution at around $-82$\,$^{\circ}$C and $-72$\,$^{\circ}$C (see Fig.~\ref{OMC1-2}), which correspond to the most habitual attitudes of the spacecraft with respect to the Sun. The average temperature has increased from $-79$\,$^{\circ}$C to $-75$\,$^{\circ}$C,
probably due to a slow degradation with time of the radiator dissipation.

\subsection{OMC data and scientific results}

Since the telemetry rate allocated to OMC does not allow to download one full image per integration, a number of around 100 CCD sections of 11$\times$11 pixels
around pre-defined objects of interest
(including the target of the main instruments) are extracted from each observation and downloaded to Earth.
The CCD sections are processed at the INTEGRAL Science Data Centre (ISDC), being corrected for bias, dark current and flat-field and being photometrically calibrated.
The processed and fully calibrated data can be accessed through the INTEGRAL archive interface at the ISDC, by using the standard ISDC query tools,
or classified by object identification from the OMC database at the Centro de Astrobiolog{\'\i}a Spanish Virtual Observatory
repository (see \citealt{SDC1} and \citealt{SDC2} for more details).\footnote{~\url{https://sdc.cab.inta-csic.es/omc/}.}
Currently, the archive contains photometric observations for almost 250\,000 sources, including around 160\,000 scientific targets and some 90\,000 calibration
sources. Out of the scientific targets, there are more than 90\,000 sources with more than 50 photometric points in the archive, including a wide variety of
optically variable sources, from individual stars to binary systems and AGN.

\begin{figure*}[ht!]
  \centering
  \includegraphics[width=0.45\linewidth]{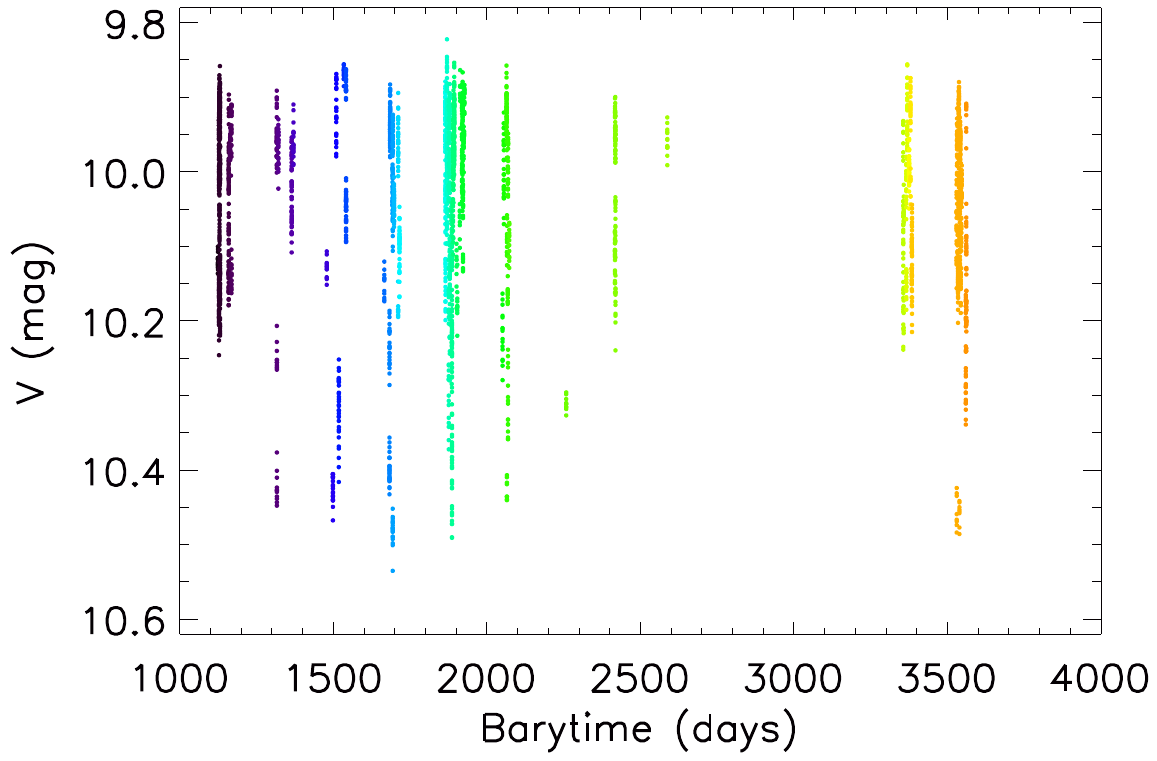}
  \includegraphics[width=0.45\linewidth]{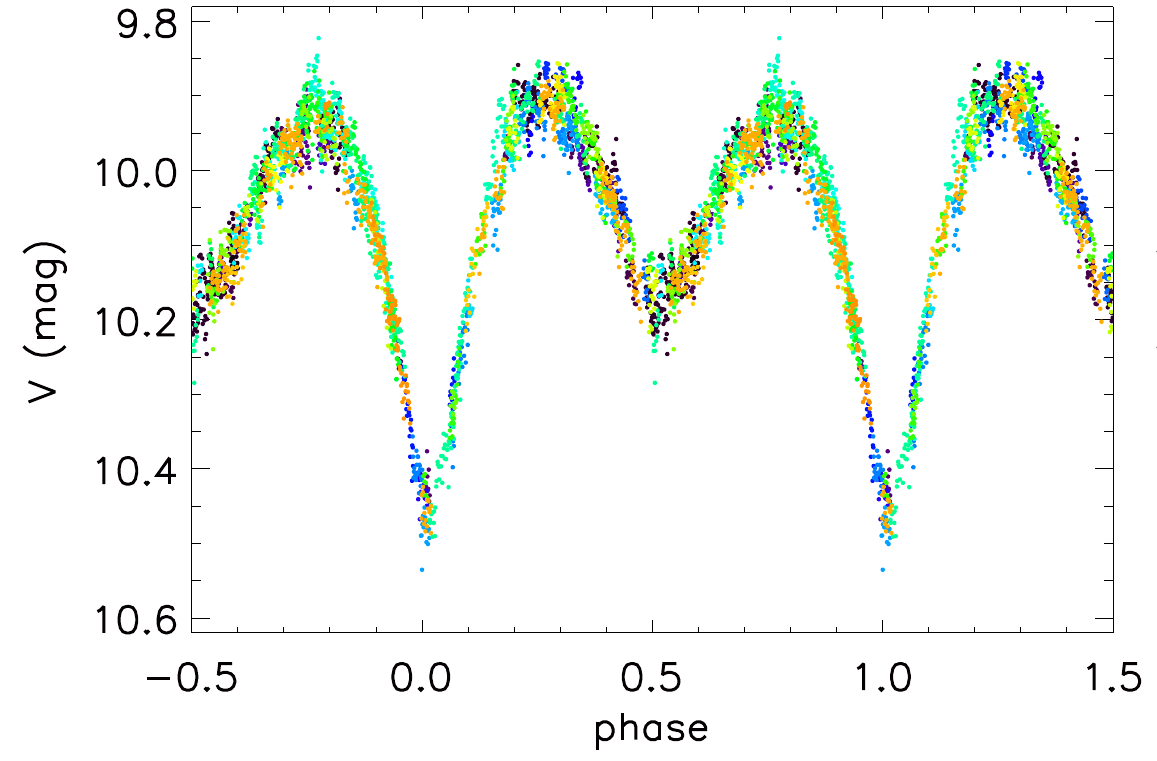}
   \caption{OMC Optical light curve over 2500 days of the eclipsing binary system IT\,Nor (IOMC 7864000021). Different colours have been used to distinguish between the different epochs, as shown in the unfolded light curve (left panel). The improvement in the determination of the orbital  period of the system ($P_{orb}=1.75414\pm0.00001$ days) with respect to previously published data, yielded a perfect phase-folding of the light curve (right panel). This is a representative example of the OMC light curves provided in \citealt{julia2012}.}
   \label{OMC2}
\end{figure*}

In this way, the first systematic analysis performed on the OMC archive light curves revealed the presence of 5263 variable sources. They are compiled in the
first catalogue of variable sources observed by OMC (\citealt{julia2012}, see, e.g., Fig.~\ref{OMC2}).
For 1337 of these variable sources, a significant periodicity could be found. The determined periods were in the range from
a few hours to some hundreds of days, with most typical values around 15\,h.
Several multi-wavelength studies have been performed making use of OMC data. Of special relevance was, for example, the monitoring with IBIS, JEM-X and OMC during the long INTEGRAL observations of the June 2015 outburst of the black-hole X-ray binary transient V404\,Cyg \citep{rodriguez2015,julia2018}.

\section{INTEGRAL Radiation Environment Monitor - IREM}
\label{sec:irem}

\subsection{IREM characteristics}

The INTEGRAL Radiation Environment Monitor \citep[IREM;][]{Hajdas2003} comes from the first series of ESA Standard Radiation Environment Monitors (SREMs) developed in
partnership between ESA, the Paul Scherrer Institut (PSI) and Contraves Space AG.
IREM is an assembly of detectors for particle spectroscopy of protons and electrons and for dosimetry measurements
within its field of view of $\pm$20$^{\circ}$. Ten identical instruments were manufactured and calibrated in the frame of the ESA SREM program.
Seven of them were launched into space. The whole batch of SREM instruments was calibrated using the same particles, energies and fluxes.
This allowed the creation of fine-tuned and individual response matrices. The tuning was necessary due to varying values of sensitive
detector area between SREMs. Additional cross-calibrations between various SREMs and IREM are conducted in space.

IREM's performance is to date characterized by very high stability. Only a few failures have occurred. They were caused by watch-dog resets
due to errors in the memory, caused by cosmic rays. Their frequency agrees well with the estimated value using single-event effect
(SEE) irradiation data. The success of the SREM program has triggered further development of new radiation monitors.

The IREM consists of two detector heads with three detectors (D1, D2 and D3). Detectors D1 and D2 have a proton and electron threshold of 20 and 1.5\,MeV,
respectively. Detector D3 has a proton and electron threshold of about 10\,MeV and 0.5\,MeV, respectively.
The relative simple configuration of IREM allows the measurement of high-energy proton fluxes with enhanced energy resolution.
The sandwich shielding between D1 and D2 effectively blocks electrons but allows passage of protons with energies greater than 39\,MeV.
This provides the pure proton signal that can be subtracted from the electron channels.
A total of 15 fixed discriminator levels are implemented to bin the energy depositions of the detected events in corresponding counters.
Ten of them are for single events, four for coincidences between D1 and D2, and one heavy ion channel. Their levels are optimized to get
the most accurate information on the spectral shape of the detected particles.

IREM functions as an autonomous radiation monitoring device. INTEGRAL's trajectory along its highly elliptical orbit allows IREM
to probe both the dynamic outer electron belt and the interplanetary environment where cosmic rays, Solar proton and electron events are encountered.
IREM's role as a payload is to support instruments on board the INTEGRAL satellite and continuously measure particle fluxes along its orbit.
It enables the payload instruments to promptly react to elevated radiation levels, in the belts and during Solar events.
Since its switch-on shortly after the launch, IREM has been operating continuously up to the present day.
With the total mission duration to date, IREM observations covered half of the very dynamic Solar cycle No.\ 23 and the whole quiet cycle No.\ 24.
The IREM data are publicly available for the wide user community; they are part of the satellite house-keeping data and made available at the INTEGRAL Science Data Centre (ISDC) in user-friendly format (\url{https://www.isdc.unige.ch/heavens/}). A database with all the SREM data in CDF format is accessible from PSI from \url{http://srem.psi.ch/cgibin/srem_data_sec.cgi}.

\subsection{IREM performance}

The IREM data set provides one of the longest stretches of radiation belt observations. The highly variable electron environment is measured during
typical outer belt crossing time of about 10\,hr. The isotropic electron intensities may reach up to 10$^7$\,cm$^{-2}$\,s$^{-1}$, even for energies above
0.5\,MeV.
For some years directly after launch, IREM also observed protons from the outer edge of the proton belt. Their intensity reached only about
10\,cm$^{-2}$\,s$^{-1}$ for energies above 20\,MeV. Due to the orbit evolution INTEGRAL's orbit moved out of this region. The IREM spectra
can be compared with contemporary belt models. Due to the high statistical accuracy of the data, the spectral characterization is possible
up to more than 5\,MeV. The limiting factor in the absolute spectral accuracy is related to the relatively narrow FOV of IREM. The exceptionally
rich, extended, and highly reliable data set is nowadays used in improved belt models and for space weather forecasting.
For instance, \citet{Metrailler2019} used IREM and XMM-Newton SREM data to model the Van Allen Belts' trapped electrons at energies ranging from 0.7 to 1.75\,MeV.
Further improvements are possible using correlations of IREM with the direction of the local magnetic field. Monitoring of the electron belts
revealed a damped oscillating behavior towards outer space. Further studies could improve our understanding of the trapping mechanics at the belts'
outer edges.

During most of INTEGRAL's orbit IREM is outside of the Earth's magnetosphere with semi-constant and rather low fluxes of cosmic rays.
As a result, IREM detected a large number of Solar events. High-energy particle fluxes from 
coronal mass ejections (CMEs) can have durations from hours to days, and protons energies up to hundreds of MeV. With the coverage of
two Solar maxima, IREM also provides a unique set of Solar energetic particle (SEP) observations (see, e.g., \citealt{Georgoulis2018}).

The most common type of radiation detected by IREM is cosmic rays. Their intensity correlates with the Solar cycle.
With IREM observations, cosmic-ray intensities and their spectral hardness as a function of time within a Solar cycle can be studied in detail.
Further cross-measurements with SREM instruments on-board other missions (e.g., Rosetta) provide comparisons of the cosmic-ray rates as a function of
the place within the Solar system (\citealt{Honig2019}).

Additional phenomena frequently detected with IREM are the so-called Forbush decreases. They are related to the propagation of Solar CMEs and
induced changes in interplanetary magnetic field structures, locally affecting cosmic-ray spectra.
Neutron monitors on Earth are typical instruments for the detection of Forbush decreases.
IREM can observe these as a small decrease in the integral flux. It provides better timing correlation with CME occurrences,
as the measurements are done in-situ. Again, combining IREM observations with data from other SREM instruments placed at
different locations in the Solar system provides useful information on the event structure and propagation (\citealt{Witasse2017}.

Even small instruments can be very helpful in the detection of extremely energetic outbursts, such as from Soft Gamma-ray Repeaters (SGRs).
For example, a flare from SGR\,1806$-$20 on 27 December 2004 saturated almost all X-ray and gamma-ray instruments in space at that time.
Only a few small detectors and radiation monitors, including IREM, were able to detect this event. Analysis of the IREM data helped
to determine the spectral shape and total energy emitted by this gigantic flare (\citealt{Hajdas2005}).

\section{INTEGRAL Science Ground Segment - SGS}
\label{sec:sgs}
The INTEGRAL Science Ground Segment (\citealt{Much2003}) is composed of the ESA-provided Mission Operations Centre (MOC) at the European Space Operations Centre (ESOC) in Darmstadt, Germany (Section~\ref{sec:moc}), and the INTEGRAL Science Operations Centre (ISOC) at the European Space Astronomy Centre (ESAC) in Madrid, Spain (Section~\ref{sec:isoc}), as well as the nationally funded INTEGRAL Science Data Centre (ISDC) at the University Geneva, Switzerland (Section~\ref{sec:isdc}). The INTEGRAL Burst Alert System, IBAS, resides at the ISDC (Sect.~\ref{sec:ibas}). Instrument Teams are also nationally funded and support the instrument-specific operations, contribute to the Offline Scientific Analysis (OSA) software, and provide the calibrations of the payload. Cooperation with MOC, ISOC, ISDC and the PIs is excellent with a collaborative forward-thinking ethos.

\begin{figure*}[ht!]
  \centering
  \includegraphics[width=0.9\linewidth, angle=0]{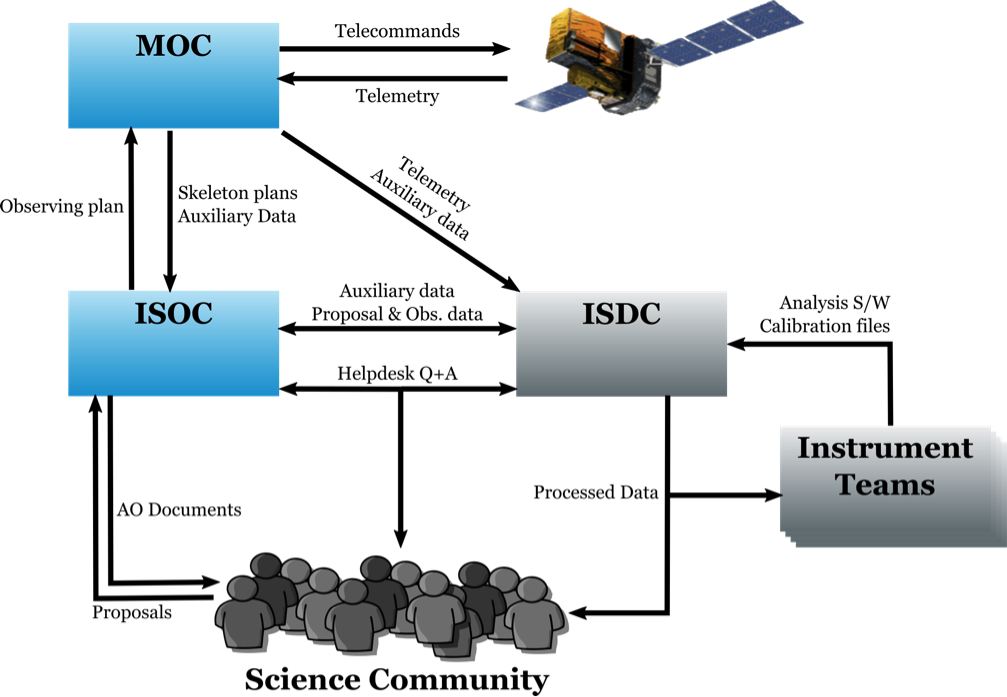}
   \caption{Schematic overview of the INTEGRAL Ground Segment.}
   \label{fig:GroundSegment}
\end{figure*}

\subsection{INTEGRAL Mission Operations Centre - MOC}
\label{sec:moc}

Preparations for INTEGRAL’s mission operations started already in 1995, with the formulation and agreement of operational requirements. Since the beginning, INTEGRAL has been controlled by MOC located at ESOC.

The mission operations planning concept has remained unchanged since launch, basically consisting of 3 steps:
\begin{enumerate}
\item Generation of a Planning Skeleton File (PSF) by MOC; defining times of orbital events such as eclipses, ground station visibility, etc..
\item The PSF is populated by ISOC with instrument configuration information and observation target attitudes to create the Planned Observation Sequence (POS).
\item Finally, the POS is processed by MOC to add all necessary engineering and attitude control commands to create an Enhanced POS (EPOS) and list of telecommands to control the spacecraft for an entire revolution.
\end{enumerate}

\begin{figure*}[ht!]
\centering
\includegraphics[width=0.95\linewidth]{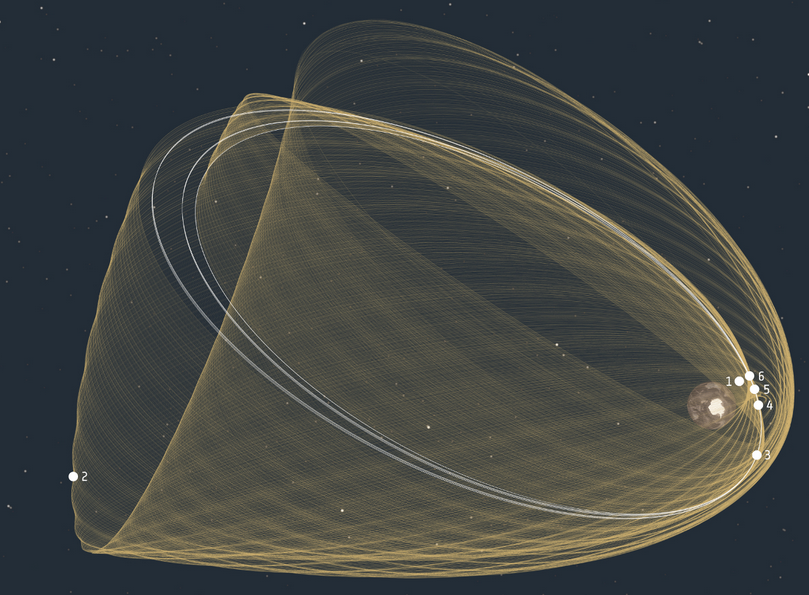}
\caption{Evolution of INTEGRAL's orbit from launch on 17 October 2002 up to October 2017. INTEGRAL travels in a geosynchronous highly eccentric orbit with high perigee in order to provide long periods of uninterrupted observation with nearly constant background and away from the radiation belts. Over time, the perigee and apogee have changed, as has the plane of the orbit. The closest approach to Earth was on 25 October 2011, at 2756\,km (1); the furthest away from Earth was on 27 October 2011, at 159967\,km (2). In 2015, four thruster burns were conducted that were carefully designed to ensure that the satellite's eventual entry into the atmosphere in 2029 will meet the ESA's guidelines for minimising space debris. These safe disposal maneuvres happened on 12 January (3), 24 January (4), 4 February (5) and 12 February (6). The orbital changes introduced during these manoeuvres are highlighted in white.}
\label{fig:orbits}
\end{figure*}

\begin{figure*}[ht!]
\centering
\includegraphics[width=0.8\linewidth]{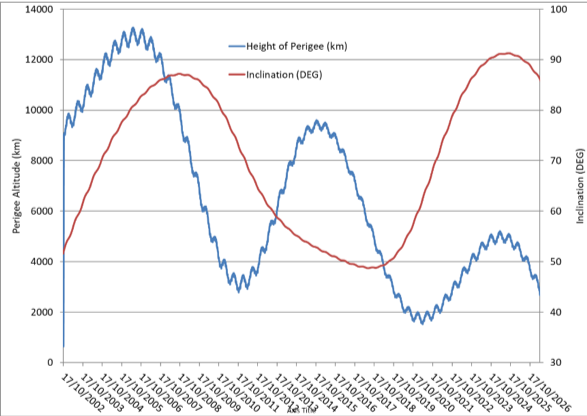}
\caption{Orbital evolution of INTEGRAL.}
\label{fig:MOC2}
\end{figure*}

\begin{figure*}[ht!]
  \centering
  \includegraphics[width=0.8\linewidth]{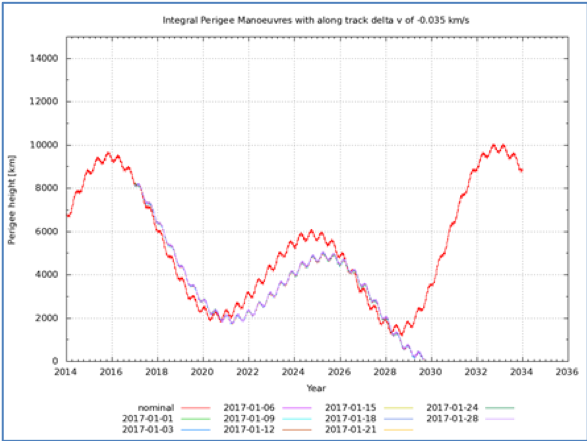}
  \caption{INTEGRAL perigee altitude evolution prediction before and after the disposal manoeuvres.}
  \label{fig:MOC10}
\end{figure*}

The orbit of INTEGRAL had to meet various demands and constraints, such as avoiding the radiation belts, maximum intervals of visibilities from the ground stations, avoiding as much as possible eclipses of the spacecraft by the Earth (see \citealt{Eismont2003}). All these requirements were to be kept within acceptable limits during the nominal and extended mission, initially foreseen to be 5 years. However, even after 18 years of operations, the approach to the orbit design is still a success.
INTEGRAL’s natural orbital evolution since launch - in particular perigee altitude and inclination (Figs.~\ref{fig:orbits} and \ref{fig:MOC2}) - has been dramatic and has been the driver for many changes in the mission operations. A further major change in the orbit was a result of the disposal manoeuvres (see Sect.~\ref{sec:Conclusions}) executed in 2015, which had two main effects:
\begin{enumerate}
\item the orbital period was reduced from 72 (3 sidereal days) to 64\,hr;
\item as a direct consequence, the orbit was no longer synchronised with the Earth’s rotation and was, therefore, no longer repeating in terms of ground station coverage.
\end{enumerate}
The influence of the manoeuvres can be seen in Figs.~\ref{fig:orbits} and \ref{fig:MOC10}.

During the early years of the mission, communications with INTEGRAL were mostly supported by ESA's Redu station in Belgium. In 2012, INTEGRAL operations were moved from Redu to Kiruna, improving satellite visibility. Kiruna remains the baseline station for INTEGRAL operations with occasional support from Weilheim (Germany), Kourou (French Guyana) and Villafranca (Spain).

\subsubsection{Mission Control System (MCS) and Flight Control Team (FCT)}

The core of the MOC operations facilities is the Mission Control System (MCS). The MCS hardware is installed in two fully redundant chains working in hot redundancy. The MCS is responsible for real-time control and monitoring of the satellite and data archiving and distribution to external partners such as ISDC, ISOC, and institutes with instruments on INTEGRAL. Continuous modernisation of the MCS allows very efficient, reliable, 24-hour operations. The current MCS will remain in place until the early 2020s when a hardware and OS upgrade will be implemented.

The FCT is responsible for executing the day-to-day activities of the mission and fulfilling the performance requirements agreed with the ESA directorate of science. The FCT consists of people with a number of speciality roles, such as Spacecraft Controllers (SPACONs), who are responsible for running the real-time operations. The FCT is supported by,  the Flight Dynamics (FD) team (responsible for all aspects of attitude and orbital control) and the Ground Facilities Operations division (responsible for the contact with INTEGRAL), amongst others. The core FCT has evolved considerably since launch according to the mission needs. Although manpower has more than halved since launch this has had no detrimental effect on performance.
In 2007 the exploitation of synergies with XMM-Newton at MOC allowed a merger and reduction of the two dedicated teams of SPACONs.
In 2018 the SPACON teams of INTEGRAL and XMM-Newton were merged with that of Gaia. For INTEGRAL the impact was very low; the higher level of on-board autonomy (compared to XMM-Newton) and  the robust and highly automated ground operations meant that science operations continued as usual.

\subsubsection{Satellite health, power and propellant}

After all these years in orbit, INTEGRAL remains in excellent health with only minor degradation. The lifetime-limiting resources should allow science operations to continue until shortly before satellite re-entry in 2029 (Sect.~\ref{sec:Conclusions}). There are no permanent failures at unit level and relatively few component failures.

Recurring anomalies can generally be addressed quickly by the SPACONs, or in some cases, the automation system, meaning that the effect on mission performance is minimal. The impact of more serious anomalies has so far been limited for longer down-times of the payload. These events are quite rare; just 8 safe modes (ESAM: Emergency Sun Acquisition Mode\footnote{This is a basic safe mode which is acquired when any operational or equipment failure threatens to put INTEGRAL in a dangerous  attitude (power or instrument constraints), or could cause further none-recoverable damage.}) and one re-initialisation each of the payload and platform power distribution unit, causing a full payload switch-off with a recovery time of more than 4 days. Occasional crashes of a Data Processing Electronics (DPE) of an instrument require a full restart of that instrument.
In July 2019 error messages indicating a software overrun on the central on-board computer were received followed by rejection of certain command classes by the satellite. The only fix was to restart the computer --- the first time in almost 17 years. The root cause is thought to be radiation damage to one of the interfacing components of the on-board computer, causing a slower response time and subsequent programme overrun. A solution was found, and since then the problem has not re-occurred.

Permanent failures are limited to:
\begin{enumerate}
\item 4 of the original 19 SPI Germanium detectors (see Sect.~\ref{sec:spi}).
\item Two sets of the SPI Front End Electronics (FEEs). Internal redundancy in the instrument means that this has no impact.
\item A number of Remote Terminal Unit (RTU) digital acquisition channels have been lost.
\item Exhaustion of hydrazine availability in April 2020 (See Sect.~\ref{sec:anomaly2020}).
\end{enumerate}

\begin{figure*}[ht!]
\centering
\includegraphics[width=0.9\linewidth]{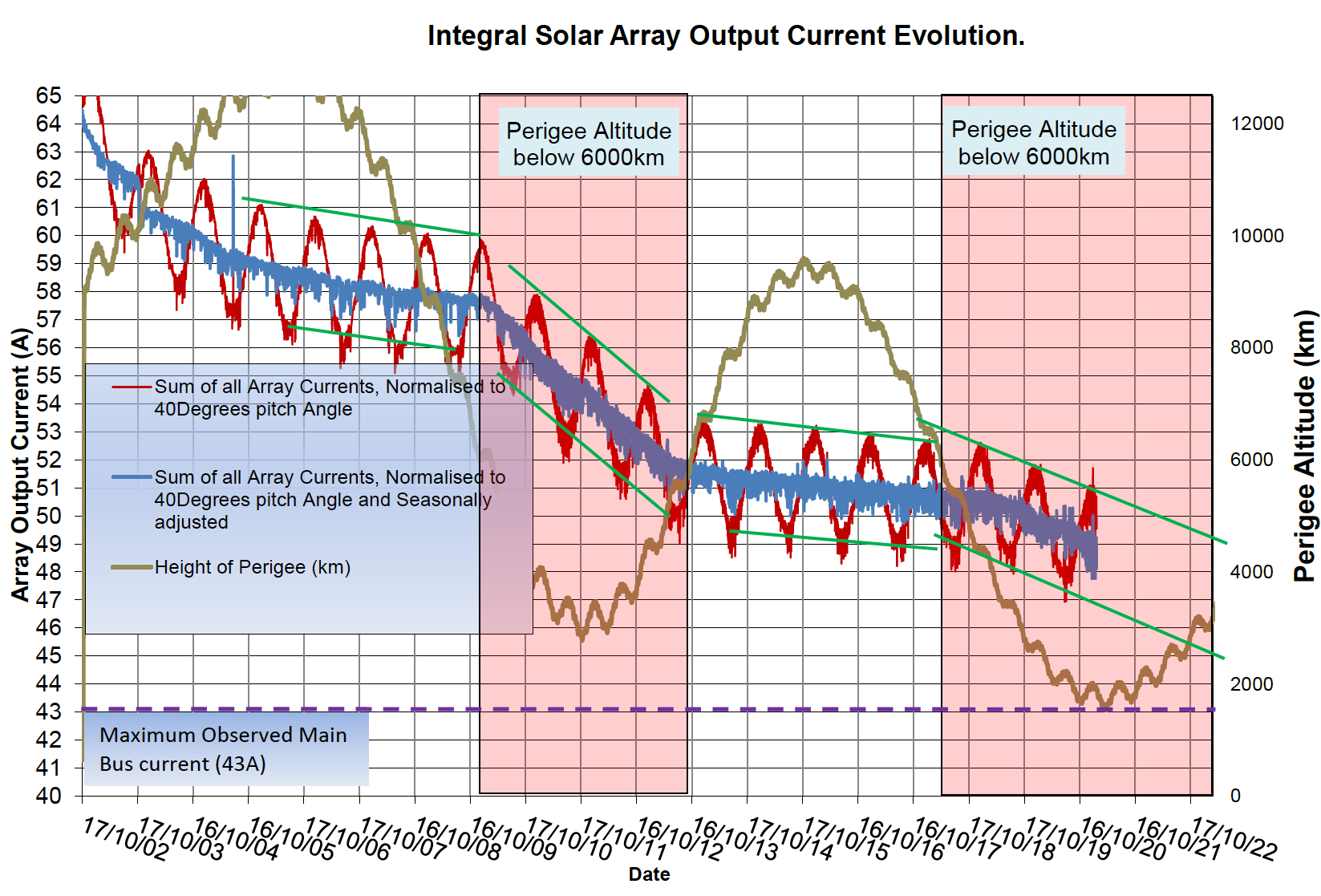}
\caption{INTEGRAL worst case Solar array output and perigee altitude evolution.}
\label{fig:MOC5}
\end{figure*}

The two critical lifetime-limiting elements are power and propellant, both elements have threatened a possible premature end to the mission. In both cases the FCT found ways to mitigate this threat. Fig.~\ref{fig:MOC5} shows the evolution of the Solar array output current since launch. It can be clearly seen that in the period when perigee altitude was below 6000\,km, exposure to trapped protons (probably) in the Van Allen belts led to an accelerated rate of degradation. Starting from early 2018, the perigee altitude has again fallen below 6000\,km and will remain there, raising concerns that power restrictions will become necessary in the early 2020s. However, the design of the satellite allows use of the batteries also in sunlight, to supplement the array output current for limited periods. Use of this mode should allow almost unrestricted science operations until beyond 2025. It can also be clearly seen that the expected increase in the rate of degradation over the last year has not occurred, raising hopes that unconstrained operations can continue even longer.

\begin{figure*}[ht!]
\centering
\includegraphics[width=0.9\linewidth]{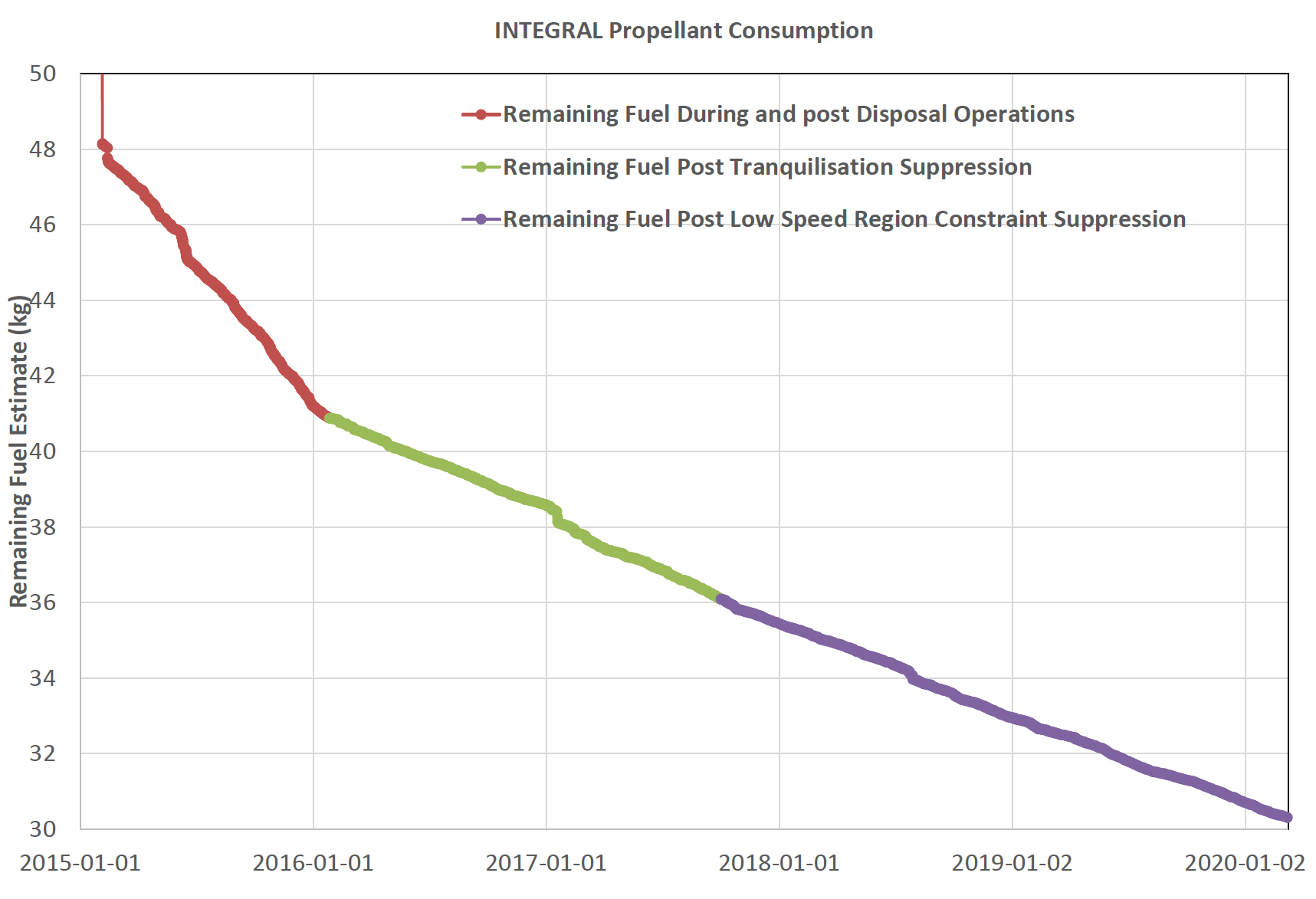}
\caption{Impact of propellant usage reduction measures since the de-orbiting manoeuvres in January 2015.}
\label{fig:MOC7}
\end{figure*}

Due to the excellent performance of the Proton launcher, INTEGRAL had a very generous propellant reserve. This changed drastically in 2015 following orbit manoeuvres to ensure safe re-entry in 2029 (see Sect.~\ref{sec:Conclusions}). To change the orbit about 50\,kg of propellant were used, corresponding to about half the remaining quantity. At the historic rate of propellant usage, INTEGRAL would have reached end of life in 2021. In 2016 and late 2017, the wheel biasing strategy was changed twice. The results of these measures can be seen in Fig.~\ref{fig:MOC7}; the consumption has been reduced from about 20\,g\,day$^{-1}$ to around 7\,g\,day$^{-1}$. The impact of the disposal manoeuvres on mission lifetime was thought to be fully mitigated.

\subsubsection{A new pointing strategy: the Z-flip}
\label{sec:anomaly2020}

On 16 May 2020 at 15:30:20 (in Revolution 2228), at the end of a reaction wheel bias (RWB), a severe under-performance of one thruster induced an ESAM (the 8th since launch). Whilst still in ESAM, the satellite unexpectedly de-pointed about 75$^{\circ}$ from the safe attitude before recovering within seven minutes. During this phase, a 5\% drop of the propellant pressure was observed. After two hours, on 17th May at 8:50 UT, the Mission Operation Control decided to actuate a fast ESAM recovery and put the spacecraft back into reaction-wheel control using the wheels at low rotation speeds. During the same day, SPI was reactivated as the spacecraft was staring in one direction with a slow drift. On 25th May, IBIS and JEM-X were also reactivated, but the normal timeline of observations was restarted only on 12th June.

\begin{figure*}[ht!]
    \centering
    \includegraphics[width=0.8\linewidth]{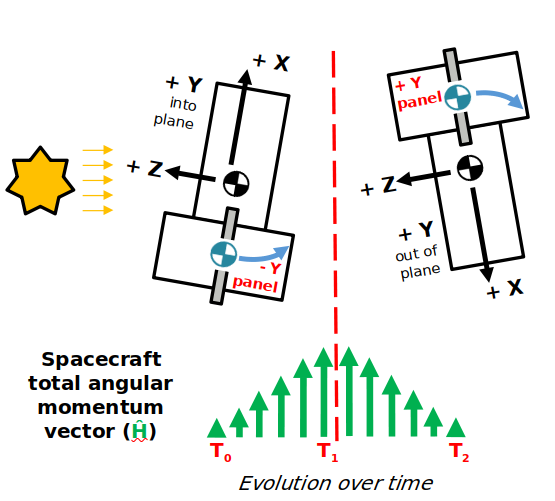}
    \caption{Illustration of the Z-flip strategy to cancel out the total angular momentum accumulated by the spacecraft, implemented in June-July 2020. Until that moment, it  was necessary to fire the thrusters every few days to unload the angular momentum accumulated in the reaction wheels (figure courtesy Dave Salt, ESA/ESOC).}
    \label{fig:z-flip}
\end{figure*}

It became gradually clear that the remaining fuel was no longer available for use, possibly due to diffusion through the membrane that separates hydrazine from nitrogen in the tanks. Indeed, each RWB maneuver performed until 20th June led to a gradual decrease of the pressure inside the propulsion system from the initial 5.5 bar to 4.2 bar on 20th June. At that point, the MOC team managed to implement a new pointing strategy in which the accumulated angular momentum is redistributed between the reaction wheels by “flipping” INTEGRAL by 180$^{\circ}$ around its Z-axis (hence 'Z-flip') using reaction wheels. In this way, the total angular momentum accumulation rate changes sign leading to a net balance over a suitable time interval, as illustrated in Fig.~\ref{fig:z-flip}. Initially, this was done within a single revolution, afterwards, it was recognized that three revolutions were sufficient. The first test was performed on 23rd June 2020 and the last RWB made on 17th July 2020.

At the time of writing, refinement of this procedure is still ongoing. Crucially, ISOC now has the challenging task of planning the satellite momentum control as well as the science activities (see Sect.~\ref{sec:rethinking}). However, it was demonstrated that this strategy does not significantly impact the usable science time while it allows potentially faster re-pointing.
On-board software updates planned at MOC will further de-constrain ISOC planning and allow greater flexibility in planning science activities.

\subsection{INTEGRAL Science Operations Centre -- ISOC}
\label{sec:isoc}
\subsubsection{Overview}

The INTEGRAL Science Operations Centre (ISOC)\footnote{\url{https://integral.esac.esa.int}} is the mission's main interface with the scientific community (see Figure~\ref{fig:GroundSegment}).
ISOC organises the yearly calls of the Announcement of Opportunity (AO) for observing proposals, performs the technical evaluation of submitted proposals, and supports the Time Allocation Committee (TAC) in the proposal evaluation process.
Based on the observations recommended by the TAC and approved by the ESA Director of Science,
ISOC takes care of long- and short-term planning of observations, including instrument handling, calibration observations, and Target of Opportunity (ToO) observations.

ISOC is responsible for the development and provision of the INTEGRAL Science Legacy Archive (ISLA), in collaboration with the ISDC and the ESAC Science Data Centre (ESDC).
ISOC maintains the mission's documentation, and supports the science community via the INTEGRAL Helpdesk, whose attendance is shared with ISDC.
These tasks are supported by a range of software tools the majority of which are developed and maintained for and by ISOC.

The development and setup of ISOC began in 1995 at the European Space Research and Technology Centre (ESTEC) in Noordwijk, The Netherlands, with a gradual buildup. ISOC was operational for the pre-launch first Announcement of Opportunity (AO-1) in 2000.
As part of the decision in November 2003 to extend INTEGRAL operations beyond the nominal lifetime of 2.2 years, it was also decided to relocate ISOC to ESAC.
The transition took place in 2004, with the timetable largely driven by the need to keep ISOC operational and able to support AO-3.

The move to ESAC saw a significant reduction in manpower, changing from $\sim$12.5~Full-time equivalents (FTE) to $\sim$9~FTE, as well as significant changes in key personnel.
Staffing remained stable over the next 10 years. In 2009 two calls for proposals each year were introduced: the first for proposals to which would be allocated observing time, and a second for proposals to which would be allocated data rights. This created a continuous and intense rhythm of preparing and handling AOs on a six-month cycle.

In 2013, strong cost pressures on the mission prompted the implementation of a cost-savings plan that led to the reduction of the available manpower at ISOC by half, from 9 to to 4.5~FTE.
This major cut in cost was made possible by reducing services and increasing efficiencies.
The measures included giving up ToO support on weekends and holidays, abandoning the second AO stage for data right proposals, reducing support for special or complicated observations, dropping INTEGRAL archive development at ESAC, reducing ISOC software support to a minimum (from 2 to 1~FTE).

Between November 2014 and April 2016 further improvements were made with efforts dedicated to streamlining procedures and workflows.
This began with a thorough review of every aspect of the science operations performed at ISOC, whose analysis resulted in a two-part plan addressing activities related to scheduling, on the one hand, and to the AO, on the other.

At the start of 2018, with a change in ISOC management and coordination, a new way of working  was introduced based on a distributed structure of Circles responsible for specific aspects relating to the phases of the yearly ISOC workflow.\footnote{For more details see Holacracy: \url{https://www.holacracy.org}.}
This was coupled with the adoption of agile development based on the SCRUM methodology, together with the modern  standards for issue tracking, documentation, sprint planning, and project management provided by Atlassian (Jira, Confluence, Bitbucket).\footnote{For more details see Atlassian software: \url{https://www.atlassian.com/software}.}
The yearly workflow is divided in four main three-month phases, each comprising three one-month sprints.
Product ownership is held by an ISOC scientist and usually rotated every 1--3 sprints depending on their scope and on the phase of the year.

\subsubsection{Calibration activities and legacy archive}

It is essential for the scientific legacy of the mission to have the best possible calibration, but almost 20 years after launch, resources are scarce and expertise has dispersed. Therefore, also in early 2018, ISOC initiated efforts to revive calibration activities with the JEM-X and ISGRI teams.
Dedicated contracts were put in place, and a strict timeline of tasks and milestones to ensure steady progress towards completion were defined and agreed upon.
Completion of this work, both at DTU Space, National Space Institute, and at the ISDC, is foreseen to be achieved in the course of 2021.

At the start of 2019, in anticipation of the end of operations at which point the responsibility for the scientific community's support and exploitation of the mission's data archive is transferred from the ISDC to ESA, the development of the INTEGRAL Science Legacy Archive (ISLA) at ESAC was kicked off.
Development is led and overseen by an Archive Scientist at ISOC, with some support from ISDC experts, working in close collaboration with the ESAC Science Data Centre (ESDC).
ISLA is the first using the most current technologies, and therefore the forerunner of the new generation of ESDC developed and maintained data archives.

 \subsubsection{Rethinking operations 18 years after launch}
\label{sec:rethinking}

\begin{figure*}[ht!]
\centering
\includegraphics[scale=.75]{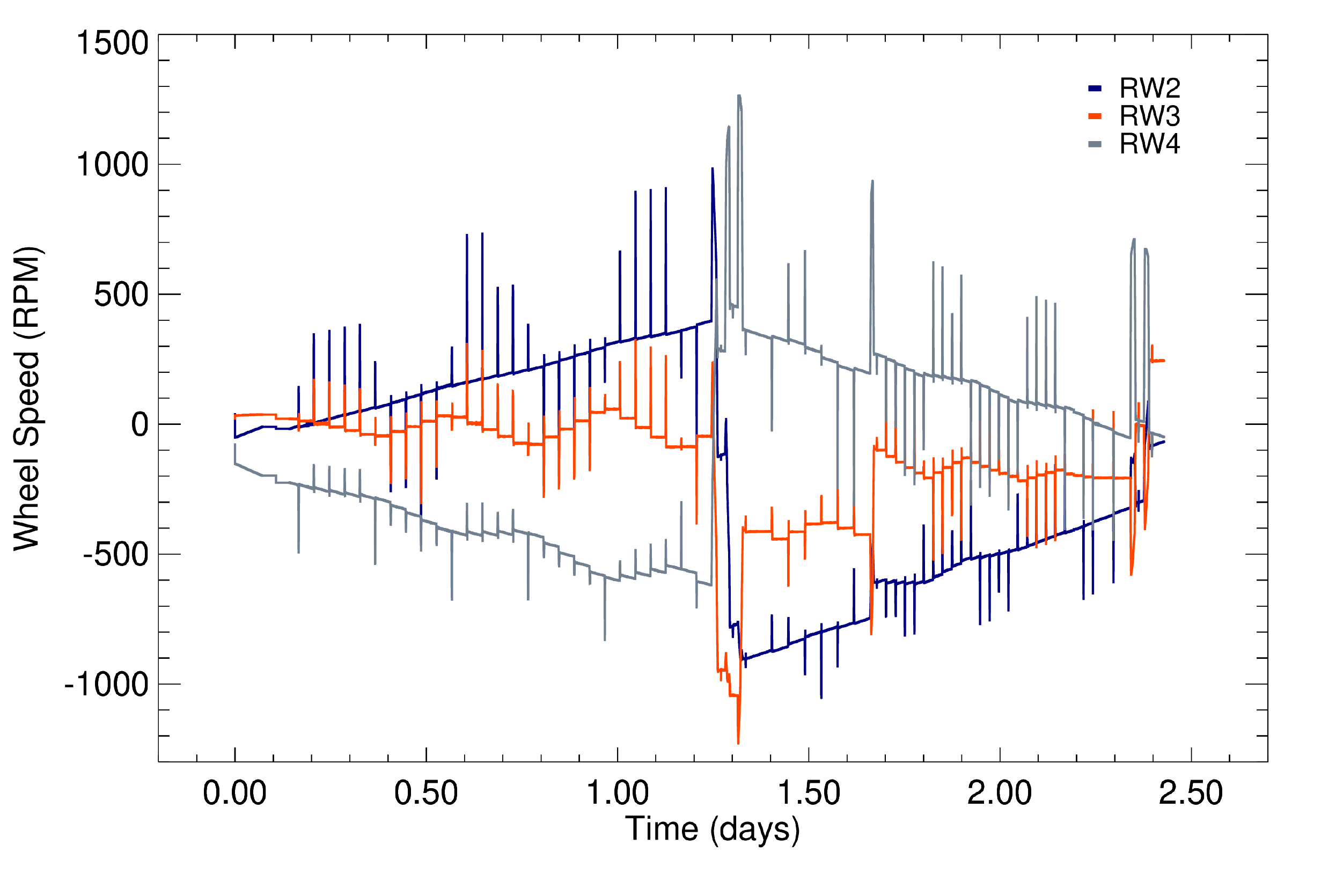}
\caption{Reaction-wheel speed profiles for Revolution 2330 (19--21 May 2020). Angular momentum is built by the Sun-radiation pressure-torque on INTEGRAL, and stored in the spacecraft reaction wheels, thus increasing their rotation speeds. Performing a 180$^{\circ}$ slew about the Sun-line at the middle of the revolution, the total angular momentum is conserved, but transferred between wheels 2 and 4. Moreover, after the slew, the torque exerted by the Sun radiation pressure is reversed, causing a decrease of the wheel speed in the second half of the revolution. A careful selection of the time of execution of the 180-degree slew allows a nearly perfect angular momentum compensation, and an angular momentum balance close to zero at the end of the revolution.}
\label{fig:ISOCz-flip}
\end{figure*}

As discussed in Sect.~\ref{sec:anomaly2020}, the automatic triggering of the 8th Emergency Sun Acquisition Mode (ESAM) on 16 May 2020 was a critical event. It marked the start of an intense period of extended meetings and technical discussions aimed at both understanding the cause of the malfunction, mitigating the possibility of a similar situation from occurring again, and finding the safest way to return to science operations. A re-thinking of satellite control without the use of thrusters and, as a consequence, also science operations, was carried out. In essence, the active control of angular momentum build-up using the thrusters was replaced by a passive control using a specific pointing strategy, i.e., the so-called 'Z-flip strategy'.
While the Z-flip strategy can be done over any number of revolutions (see Sect.~\ref{sec:anomaly2020}) as long as the total angular momentum remains below a specific safety threshold, it is most efficient to implement on a revolution-basis. This provides the lowest angular momentum buildup, and therefore the greatest flexibility in responding to, e.g., ToO requests requiring an unplanned re-pointing. An example of a wheel-speed profile over 1 revolution is given in Fig.~\ref{fig:ISOCz-flip}.  

\subsection{INTEGRAL Science Data Centre - ISDC}
\label{sec:isdc}

The INTEGRAL Science Data Centre \citep[ISDC]{Courvoisier2003}, nowadays the ISDC Data Centre for Astrophysics, receives the telemetry stream from the Mission Operation Centre (MOC; Sect.~\ref{sec:moc}),
it decodes it and distributes the INTEGRAL data to the general community in a FITS-based format through an online archive.
The ISDC was established in 1995 by a consortium of 12 institutes. The ISDC is affiliated to the Geneva Observatory,
itself part of the astronomy department of the University of Geneva.
The staff increased from very few in 1995 to some 40 in June 2003, and then slowly decreased thereafter.
The current team works in Ecogia, a rural settlement near Versoix, some 8\,km from Geneva.

The ISDC developed the core libraries of the analysis software and performs partial development, packaging, compilation, and testing of tools that constitute the Offline Scientific Analysis (OSA).
The OSA package allows each user to produce images, spectra, and light curves of celestial sources starting from the distributed data.
The ISDC maintains an up-to-date version of the instrument calibration files as part of the overall archive and responds to queries of users related to data,
analysis and software issues through a Help-desk, shared with the INTEGRAL Science Operations Centre (ISOC; Sect.~\ref{sec:isoc}).

The production of INTEGRAL data is done on a multi-level basis: from level zero to level one, the telemetry is decoded and
stored in FITS files. From level one to two, telemetry is combined with auxiliary files transmitted from MOC, and produced on the
basis of the instrument characteristics. This processing is performed on both the Near-Real Time (NRT) telemetry and the
Consolidated (CONS) telemetry. The NRT data reaches the ISDC with a latency of a few hours after the observation, and these are available to the users soon
thereafter. The CONS data are provided about ten days later by the MOC, after performing checks and possible recovery of telemetry packets.
Only for the NRT data, level two to level three data are produced.
This is done to allow a quick-look analysis of the data, in order to detect new and unexpected sources, as well as to monitor the instruments.
Owing to this activity, many new sources have been quickly found and communicated via telegrams and papers.

All calibrations and instrument characteristics that are
needed in the different steps of the analysis are stored in a set
of files. For each file the epoch of its creation as well as the
epochs for which it is valid are provided. The files are updated as knowledge of the instruments increases and as their
characteristics vary with time. The instrument characteristics files are accessed through the
ISDC archive\footnote{~\url{https://www.isdc.unige.ch/integral/archive}.}. They are organised such that the analysis software
can access them without manual intervention.
The archive contains after 18 years of mission about 20 Terabytes of data, all accessible via
\texttt{ftp} or \texttt{rsync} protocols and searchable with a dedicated version
of \texttt{w3browse}.

The OSA software can be freely downloaded and is fully documented\footnote{~\url{https://www.isdc.unige.ch/integral/analysis}}.
It is written in a modular style, with pipelines combining the single
executables which need to be run in sequence.
The pipeline processes each science window (ScW)\footnote{A Science Window (ScW) is a continuous time interval during which all data acquired by the INTEGRAL instruments result from a specific S/C attitude orientation state.} separately, and
combines results to produce mosaicked images, or average spectra, as well as long-term light curves of individual sources.
An essential contribution for single executables is given by the instrument teams, who have the best knowledge of the instrumental details.
ISDC has combined these elements and compiled handbooks, which are updated regularly.
The long history of OSA releases is summarized in Table~\ref{tab:osareleases}.
An online version of OSA is in advanced development\footnote{~\url{https://www.astro.unige.ch/cdci/astrooda_}}.
This web interface facilitates the access of high-level products also to non-experts of
INTEGRAL analysis without the need of installing software or downloading the archive.

\begin{table}
    \centering
    \label{tab:osareleases}
    \caption{Release dates of the Offline Scientific Analysis (OSA) by the ISDC.}
    \begin{tabular}{cc}
        \hline
        \hline
		\multicolumn{1}{c}{OSA} & \multicolumn{1}{c}{Release} \\
		\multicolumn{1}{c}{Version} & \multicolumn{1}{c}{Date} \\
        \hline
        1.0 & 14 April 2003 \\
        1.1 & 27 May 2003 \\
        2.0 & 21 July 2003 \\
        3.0 & 9 December 2003 \\
        4.0 & 18 June 2004 \\
        4.1 & 9 September 2004 \\
        4.2 & 15 December 2004 \\
        5.0 & 1 July 2005 \\
        5.1 & 24 November 2005 \\
        6.0 & 15 January 2007 \\
        7.0 & 28 September 2007 \\
        8.0 & 31 August 2009 \\
        9.0 & 22 March 2010 \\
        10.0 & 18 September 2012 \\
        10.1 & 4 September 2014 \\
        10.2 & 10 December 2015 \\
        11.0 & 19 October 2018 \\
        11.1 & 15 September 2020 \\
        \hline
    \end{tabular}
\end{table}

The gamma-ray burst (GRB) alert system IBAS is run from the ISDC; the IBAS is described in the next  Sect.~\ref{sec:ibas}.
In recent years, the ISDC has developed a program for targeted search of impulsive signals from all INTEGRAL instruments to detect even the
weakest GRBs linked to GWs, very-high energy neutrinos, or other relevant celestial events.
The dissemination of these results is currently part of the core scientific activity of ISDC (see \citealt{Ferrigno2021}).

\subsection{INTEGRAL Burst Alert System - IBAS}
\label{sec:ibas}

Although INTEGRAL was not specifically optimized for the study of gamma-ray bursts (GRBs),
it was immediately realized that its unprecedented imaging capabilities in the soft
gamma-ray range, offered the chance  to get accurate localizations for these events. The expected rate of bursts in the IBIS field of view of 30$\times30$
deg$^2$ was in the range $\sim$10--20 per year \citep{pedersen97}, clearly smaller than the rate  of GRB detections by instruments available at that time such
as the BATSE/\-CGRO, but with the advantage of immediate localization at the level of only a few arcminutes. The continuous telemetry link was another asset of the
INTEGRAL mission, since it gave the possibility of distributing the GRB triggers and localizations worldwide in near-real time.

It was thus decided to implement the INTEGRAL Burst Alert System (IBAS, \citealt{ibas}) for the detection and localization of GRBs and other short duration
transients using ground-based software, with the requirement that all the relevant telemetry data  had to be available for this task at the ISDC with the
shortest possible delay after their acquisition by the satellite.
During the INTEGRAL performance and verification phase two bursts were detected, in good agreement with the predictions:
GRB021115 \citep{malaguti2003} and GRB021219 whose coordinates were distributed 10 seconds after the start of the
burst \citep{Mereghetti2003}. Thanks to IBAS, INTEGRAL has been the first mission to distribute in real time the positions of
GRBs with arcminute accuracy\footnote{Note that the Neil Gehrels Swift Observatory, dedicated to rapid follow-up of GRBs, started routine operations in 2005.}.

IBAS was implemented with a flexible multi-thread architecture that allows different triggering algorithms to
operate in parallel. The GRB localization is based on data from the IBIS/ISGRI detector (Sect.~\ref{sec:ibis}) that are used in two ways.
To investigate the shortest timescale, a trigger based on the examination of the overall count rate in different energy ranges and time bins is used. When a
statistically significant excess with respect to a running estimate of the background is found, an imaging analysis is carried out to confirm the event. The
other triggering algorithm, used for timescales longer than 5\,s, is based on a continuous comparison of images and it allows the detection of
slowly varying sources also in the presence of background variations and/or other variable sources in the field of view.

IBAS also makes use of the data obtained with the anti-coincidence shield (ACS) of SPI (Sect.~\ref{sec:spiacs}). Although no positional and spectral
information is provided by these data (light curves binned at 50\,ms in a single channel at energy above $\sim$75\,keV), the large effective area makes
this instrument the most sensitive omni-directional detector currently available \citep{Savchenko2012}, with about 200 GRBs detected each year.
IBAS automatically produces and distributes SPI ACS light curves of the triggered events that are routinely used for GRB localization based on triangulation
with other satellites of the IPN network.

Up to February 2021, 136 GRBs have been detected in the IBIS field of view\footnote{ ~\url{http://ibas.iasf-milano.inaf.it/IBAS\_Results.html}.}.
For 75\%\ of them the position, with typical uncertainty of 1--2 arcmin, was publicly distributed in near-real time, i.e., within a few seconds.
Most of the slower localizations are confined to ``sub-threshold'' bursts that required human intervention to be confirmed.
Since February 2011, the positions of these lower significance events are also distributed in real time.

\begin{figure*}[ht!]
\centering
\includegraphics[scale=.45]{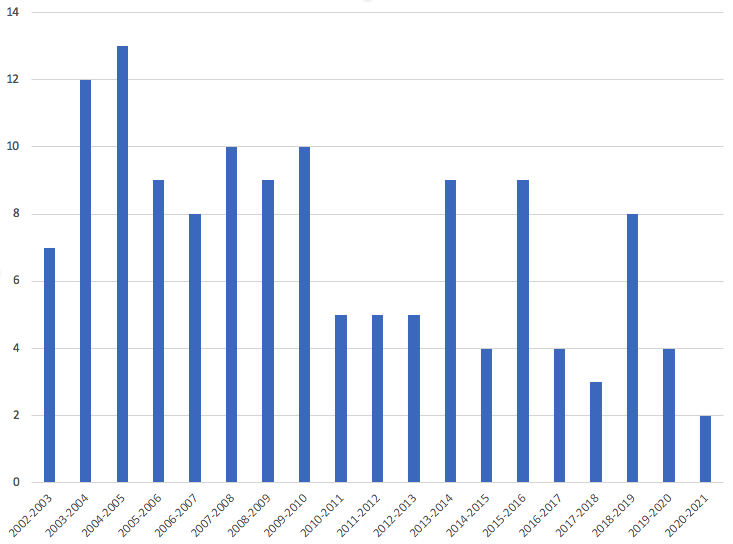}
\caption{Number of GRBs per year localized by IBAS.}
\label{fig:grbyr}
\end{figure*}

\begin{figure*}[ht!]
\centering
\includegraphics[width=0.8\linewidth]{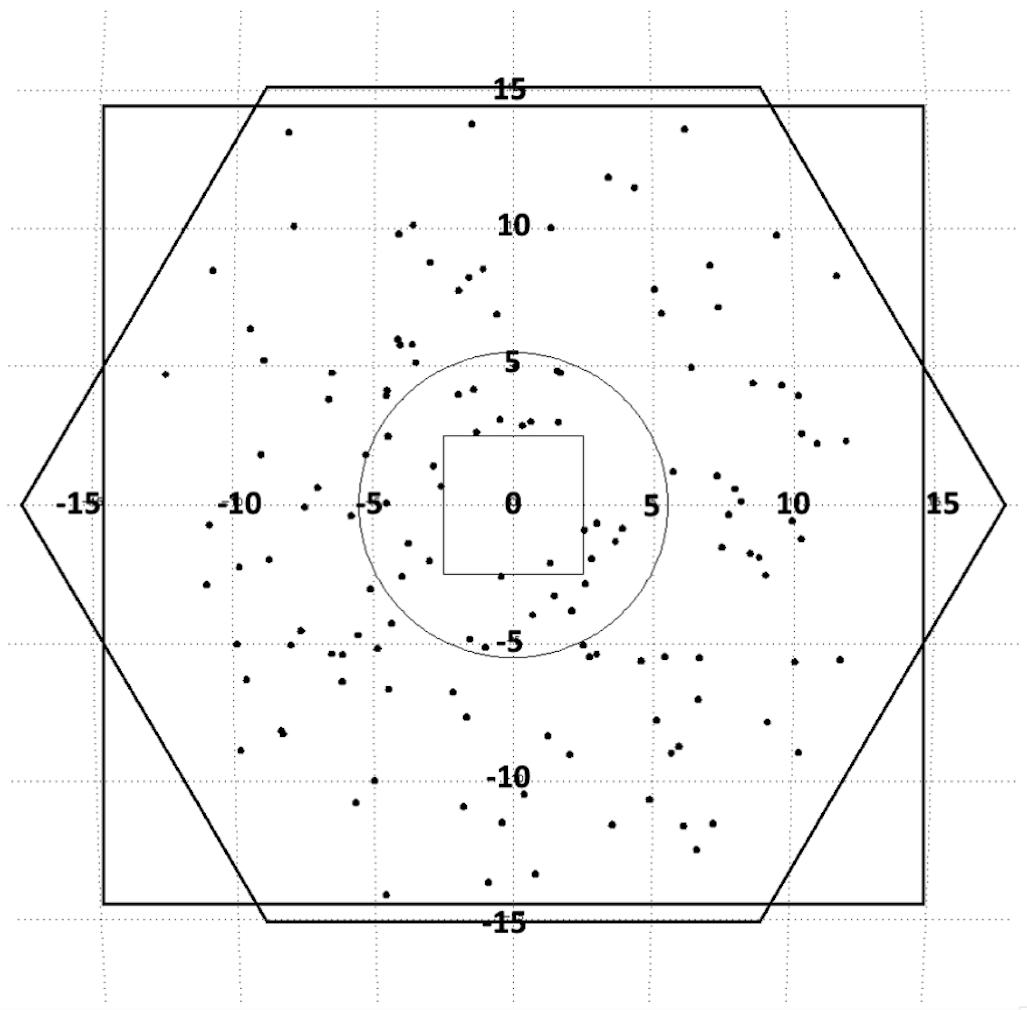}
\caption{Distribution of GRBs in the fields of view of the INTEGRAL instruments: small square: OMC, circle: JEM-X, hexagon: SPI, big square: IBIS.}
\label{fig:fov}
\end{figure*}

Figure \ref{fig:grbyr} shows that the average rate of bursts localized by IBAS is about 7.5 per year. The rate was higher during the first decade.
The subsequent decrease is possibly caused by the gradual increase over time of the ISGRI low-energy threshold (see Sect.~\ref{sec:ibis}).
The distribution of the GRBs in the field-of-view of the INTEGRAL instruments (up to February 2021) is shown in Fig.~\ref{fig:fov}, where it can be seen that bursts
can be efficiently detected and well-localized, even at very large off-axis angles.

\section{Conclusions and outlook}
\label{sec:Conclusions}

INTEGRAL will remain the only observatory providing wide-band (3\,keV--10\,MeV) imaging and gamma-ray spectroscopic capabilities to the astronomical community for the coming years.
The mission’s unique capabilities, high-resolution gamma-ray spectroscopy, polarization, large field-of-view, unparalleled imaging dynamic range, and Target-of-Opportunity programme, provide outstanding evidence that the mission will continue to provide world-class science. 
The Flight Control Team is stable with a very high level of expertise; sharing of manpower with other missions has allowed the INTEGRAL MOC to keep a large pool of expertise on hand at moderate cost. The overall Ground Segment is stable with modern reliable systems and a clear evolution path for the next decade. The Kiruna ground station visibility of INTEGRAL is sufficient to allow full science return, the performance of the antenna is excellent, and it will remain available to INTEGRAL for the foreseeable future.

The satellite health is good, except for the Reaction Control System anomaly (following Emergency Safe Attitude Mode, ESAM, \#8) experienced in May 2020. The evolution of Solar-array degradation is better than foreseen. Since the anomaly in May 2020, the INTEGRAL propulsion system has only very limited capacity for actuation, and the available thruster impulse is very low. The present system capabilities and associated risks are assessed as follows:  INTEGRAL is expected to be capable of surviving a safe-mode entry and a limited period of control depending on dynamic conditions (attitude and rates) at safe mode entry and on the time spent in safe mode, but it is doubtful if more than one or two safe modes can be accommodated. The available thruster impulse together with the extremely uncertain performance means that it is unlikely that any requested collision avoidance manoeuvre (CAM) can be completely fulfilled, nevertheless
it should be possible to influence the spacecraft orbit to reduce any risk. Following any thruster activation (safe mode or CAM) it is much less likely that any subsequent safe mode or CAM will execute correctly, and certainly not within a limited time period (about 1 month). In the 18-years mission to date we have experienced just 8 safe modes, i.e., less than 1 every 2 years.
Finally, it is noted that limited-thruster control on INTEGRAL is infinitely better than no control, given the fact that the spacecraft will remain in orbit for about 8 more years, including crossings of the Sun-synchronous and geostationary orbit belts.

The operational concept (which was stable before) had to be adjusted due to the anomaly, to handle angular-momentum control without the Reaction Wheel Bias (RWB) and thruster usage; since then (mid-July 2020) the spacecraft is thus operated without thrusters, i.e., fuel consumption is null. Hence, out of the anomaly was born an entirely new and highly efficient method that combines pointing strategy --- and thus science operations --- together with satellite control --- and thus mission operations --- into a model that may inspire future missions.
In July 2020, the INTEGRAL Users Group (IUG) wrote a letter to the ESA Director of Science in order to “express their gratitude to the MOC, ISOC and ESTEC teams for their highly collaborative approach to resolving the recent [anomaly] issues” and to acknowledge “the amazingly elegant solution [that] has been implemented to allow recovery of [at least] 90--95\% operational efficiency, while ensuring the health and continued safe operation of the mission”.

Under normal space weather and orbit conditions, the instruments should continue to collect valuable data in INTEGRAL's unique energy range of space astronomy, and continue to provide science performance of the same quality until the end of the mission.

The final act of INTEGRAL will be re-entry into the Earth’s atmosphere and destruction in February 2029. By this time, the solar arrays will be degraded to such an extent that science cannot be done anymore. Thus re-entry was deliberately engineered to prevent INTEGRAL from becoming long-lasting space debris. Therefore, in early 2015 a series of manoeuvres adjusted the orbit of INTEGRAL such that its long-term evolution would lead to the re-entry (see Fig.~\ref{fig:MOC10}). This disposal is compliant with the ESA space-debris policy and is secured even if INTEGRAL becomes no longer operable beforehand. Re-entry will occur in sparsely populated areas, south of a latitude of $-$45$^{\circ}$.

\section{Acknowledgements}

Based on observations with INTEGRAL, an ESA project with instruments and science data centre funded by ESA member states (especially the PI countries: Denmark, France, Germany, Italy, Switzerland, Spain) and with the participation of Russia and the USA.
The OMC team has been funded by different Spanish grants, including Spanish State Research Agency grants PID2019--107061GB--C61
and MDM--2017-0737 (Unidad de Excelencia Mar{\'\i}a de Maeztu -- CAB). The INTEGRAL French teams acknowledge partial funding from the French Space Agency (CNES).
The Danish JEM-X gratefully acknowledges support from the Danish PRODEX delegation through contract C90057.
The INTEGRAL Italian team  acknowledges support form the Italian Space Agency, ASI, along these years via different agreements, last of which is 2019–35-HH.0. LH acknowledges support from SFI under grant 19/FFP/6777 and the EU AHEAD2020 project (grant agreement 871158). TS is supported by the German Research Society (DFG-Forschungsstipendium SI 2502/1-1 \&\ SI 2502/3-1).

"If I have seen further, it is by standing upon the shoulders of giants" (Sir Isaac Newton in a letter to written to Robert Hooke in February 1675). General appraisal goes to the many colleagues (consult, e.g., the author lists of the special Astronomy \& Astrophysics letters issue in 2003 on: "First science with INTEGRAL": see \url{https://www.aanda.org/component/toc/?task=topic&id=652}) who conceived, built, launched and operated INTEGRAL in the beginning; without them INTEGRAL would not be where it is now.

Since launch, unfortunately, various people who contributed to the success of INTEGRAL one way or the other have left us. They are, in chronological order, Mike Revnivtsev (2016, former INTEGRAL Users' Group member), Neil Gehrels (2017, former INTEGRAL Mission Scientist), Lars Hansson (2017, former INTEGRAL Science Operations Manager), Giovanni 'Nanni' Bignami (2017, former chair of the ESA Science Evaluation Committee that selected INTEGRAL in 1995) Giorgio Palumbo (2018; former INTEGRAL Mission Scientist), and Pierre Mandrou (2019; former INTEGRAL/SPI instrument scientist).

\bibliography{mybibfile,special}

\end{document}

%% file: aa_macros.tex
\def\arxivprefixesep{:}

\newcommand{\eprint}[2][]{%
{\tt\if!#1!#2\else#1\arxivprefixesep\ignorespaces#2\fi}%
}